\documentclass[twocolumn]{aastex62}
\pdfoutput=1 
\usepackage[maxfloats=40]{morefloats}
\usepackage[T1]{fontenc}
\usepackage{apjfonts}
\usepackage{color}
\usepackage{catoptions}
\usepackage{booktabs}
\usepackage{amsmath}
\usepackage[all]{hypcap}
\newcommand{\Rom}[1]{\uppercase\expandafter{\romannumeral #1\relax}}

\begin{document}
\makeatletter

\renewcommand{\sectionautorefname}{Section\negthinspace}
\renewcommand{\subsectionautorefname}{Section\negthinspace}
\renewcommand{\subsubsectionautorefname}{Section\negthinspace}

\title{Polarimetric and photometric investigation of a dark globule LDN\,1225: distance, extinction law, and magnetic fields}

\author{Chakali Eswaraiah}
\affil{Institute of Astronomy, National Tsing Hua University (NTHU), 101, Section 2, Kuang-Fu Road, Hsinchu 30013, Taiwan, R.O.C}

\affil{National Astronomical Observatories, Chinese Academy of Sciences, Datun Road, Chaoyang District, Beijing 100101, People's Republic of China}
\affil{CAS Key Laboratory of FAST, NAOC, Chinese Academy of Sciences, Beijing 100101, People's Republic of China}
\affil{Indian Institute of Astrophysics, Koramangala 2nd Block, Bengaluru 560 034, India}

\author{Shih-Ping Lai}
\affil{Institute of Astronomy, National Tsing Hua University (NTHU), 101, Section 2, Kuang-Fu Road, Hsinchu 30013, Taiwan, R.O.C}

\author{Yuehui Ma}
\affil{Purple Mountain Observatory and Key Laboratory of Radio Astronomy, Chinese Academy of Sciences, 2 West Beijing Road, Nanjing 210008, People's Republic of China}
\affil{University of Chinese Academy of Sciences, 19A Yuquan Road, Shijingshan District, Beijing 100049, People's Repulic of China.}

\author{Anil K. Pandey}
\affil{Aryabhatta Research Institute of Observational Sciences (ARIES), Manora-peak, Nainital, Uttarakhand-state, 263002, India}

\author{Jessy Jose}
\affil{Indian Institute of Science Education and Research Tirupati, Rami Reddy Nagar, Karakambadi Road, Mangalam (P.O.), Tirupati 517 507, India.}

\author{Zhiwei Chen}
\affil{Purple Mountain Observatory, Chinese Academy of Sciences, 8 Yuanhua Road, 210034, Nanjing, China}

\author{Manash R. Samal}
\affil{Physical Research Laboratory (PRL), Navrangpura, Ahmedabad 380 009, Gujarat, India}

\author{Jia-Wei Wang}
\affil{Institute of Astronomy, National Tsing Hua University (NTHU), 101, Section 2, Kuang-Fu Road, Hsinchu 30013, Taiwan, R.O.C}

\author{Saurabh Sharma}
\affil{Aryabhatta Research Institute of Observational Sciences (ARIES), Manora-peak, Nainital, Uttarakhand-state, 263002, India}

\author{D. K. Ojha}
\affil{Department of Astronomy and Astrophysics, Tata Institute of Fundamental Research, Homi Bhabha Road, Mumbai 400 005, India}

\correspondingauthor{Eswaraiah Chakali}
\email{eswaraiahc@nao.cas.cn, eswaraiahc@outlook.com}

\begin{abstract}
We present the results based on the optical $R$-band polarization observations of 280 stars distributed towards the dark globule LDN\,1225. 
{\it Gaia} data release 2 parallaxes along with the polarization data of $\sim$200 stars have been used to 
	(a) constrain the distance of LDN\,1225 as 830$\pm$83~pc, 
	(b) determine the contribution of interstellar polarization (ISP), and (c) 
	characterize the dust properties and delineate the magnetic field (B-field) morphology of LDN\,1225. 
	We find that B-fields are more organized and exhibit a small dispersion of 12$\degr$. Using the $^{12}$CO molecular line data from the Purple Mountain Observatory (PMO), along with the column density, dispersion in B-fields, we estimate B-field strength to be $\sim$56\,$\pm$\,10\,$\mu$G, magnetic to turbulence pressure to be $\sim$3\,$\pm$\,2, and the mass-to-magnetic flux ratio (in units of critical value) to be~$<$\,1. These results indicate the dominant role of B-fields in comparison to turbulence and gravity in rendering the cloud support.
B-fields are aligned parallel to the low-density parts (traced by $^{12}$CO map) of the cloud, in contrast they are neither parallel nor perpendicular to the high-density core structures (traced by $^{13}$CO and C$^{18}$O maps). 
LDN\,1225 hosts two 70\,$\mu$m sources which seem to be of low-mass Class 0 sources. 
The total-to-selective extinction derived using optical and near-infrared photometric data is found to be 
anomalous ($R_{V}$~$=$~3.4), suggesting dust grain growth in LDN\,1225. Polarization efficiency of dust grains 
follows a power-law index of $-$0.7 inferring that optical polarimetry traces B-fields in the outer parts of the cloud. 
\end{abstract}

\keywords{Polarization - (ISM:) dust, extinction - ISM: clouds, magnetic fields - local interstellar matter: individual: LDN 1225}

\section{Introduction}\label{sec:introd}

The small, compact, and isolated dark globules known as ``Bok globules" \citep{BokReilly1947}, are the potential precursors to the protostars. Initially, \citet{Barnard1927} had prepared a list of such dark regions in the sky and later \citet{Lynds1962} published a catalogue with a larger number of such dark objects. Furthermore, \citet{ClemensBarvainis1988} have compiled a list of 248 small (mean size $\sim4\arcmin$) and nearby (distance $<$\,1 kpc) molecular clouds. 
Subsequent studies have shown that these clouds exhibit signs of star formation such as collimated molecular outflows, 
compact infrared sources, and very low luminous objects 
(VELLOs) \citep{Reipurth1983,FrerkingLanger1982,Neckeletal1985,Vrbaetal1986,NeckelStaude1990,Richeretal2000,Stecklumetal2007,Reipurth2008}. 
The characteristics of new born stars will depend on the physical conditions of the molecular cloud core prior to the onset of gravitational collapse. 

Turbulence in the dark globules is sub-or trans-Alfv\'{e}nic \citep{Heyeretal2008,Francoetal2010} and the cores 
embedded in them are characterized with subsonic turbulence \citep{MyersBenson1983,Goodmanetal1998} such that their 
effect is inadequate to counteract the gravitational collapse.
Therefore, magnetic fields (B-fields) are proposed to play a crucial role in regulating 
isolated low-mass star formation by controlling the stability and contraction of the cores in these clouds \citep{Mouschovias1976,Shu1977,Shuetal1987}.
Moreover, it has been shown that B-fields are indeed important in regulating shape of the cloud fragments, guiding accretion flows, directing the outflows, and collimating the jets of the protostars \citep{Vallee2002,Sugitanietal2010,Girartetal2006,Pudritzetal2007,MckeeOstriker2007,Gallietal2009}.
Here we study the morphological correlations between the cloud and B-fields, 
and investigate the relative importance of B-fields to turbulence and gravity in LDN\,1225 \citep[eg.,][]{MyersGoodman1991,Wardthompsonetal2000,Eswaraiahetal2013,Bertrangetal2014,Kandorietal2017,JorqueraBertrang2018}. 

The dark globule LDN\,1225 (or CB242 or H699 P16) with central coordinates of RA (J2000)~$=$~23$^{h}$11$^{m}$58$^{s}$ 
and Dec (2000)~$=$~$+$61$\degr$39$\arcmin$00$\arcsec$ (or $l$~$=$~111$\fdg$41, $b$~$=$~$+$01$\fdg$02) 
is located towards the Cepheus OB3 cloud complex (hereafter CepOB3). 
Based on the derived extinction and distance values of the stars projected 
toward the cloud, \citet{Maheswaretal2006} have bracketed the distance of LDN\,1225 as 400$\pm$80~pc. 
Nonetheless, our study suggests that, being kinematically associated with CepOB3, LDN\,1225 is located at 830$\pm$83~pc. \
A catalog of dust clouds in the Galaxy by \citet{DutraBica2002} indicates that the spatial extent of LDN\,1225 is $8\arcmin\times4\arcmin$, opacity class as 4 (1 being more lightest and 6 being more darkest; \citealt{Lynds1962}), and the local standard of rest velocity (V$_{LSR}$) based on CO data as $-$10.9 km s$^{-1}$ \citep{ClemensBarvainis1988}. Opacity class 4 \citep{Lynds1962} is similar to the density class B (A being the most dense and C being the least dense; \citealt{Hartleyetal1986}). Therefore, according to the above opacity classes, LDN\,1225 is an intermediate dense dark globule. 

In this work, our goal is to investigate 
whether B-fields support, relative to turbulence and gravity, is important to the formation and evolution of the dark globule LDN\,1225. 
For this purpose, the B-field strength and its pressure, turbulent pressure, ratio of ordered to turbulent component of B-fields, the mass-to-magnetic flux ratio in units of critical value, etc, have been estimated using $R$-band polarization and CO molecular lines data. 
Polarimetric data have been used in combination with {\it Gaia} data release 2 \citep{Gaiacollaboration2018b} parallaxes to find the cloud distance, to estimate foreground polarization contribution, and to determine the dust properties and B-fields in LDN\,1225. 
Based on the distances, kinematic information, and the coherent B-fields at small and large scales, the membership of LDN\,1225 to the CepOB3 is discussed. 
The possible correlation between multiple components of polarizations and those of CO gas is discussed. 
We discuss the properties of two 70\,$\mu$m sources and their association with LDN\,1225. 
Optical photometric data have been used to characterize the extinction law in the foreground and cloud mediums. 
Furthermore, dust polarization efficiency of the dust grains in the dark globule LDN\,1225 is studied.  

Outline of this paper is as follows. Section \ref{sec:obs_and_dataredu} describes the observation and 
data reduction of various observed data along with the archival data sets.
Analyses and results are presented in section \ref{sec:analyses_and_results}. 
Discussion based on our results is given in section \ref{sec:discuss}. Summary and conclusions of this work are mentioned in section \ref{sec:summary_conclusions}.

\section{Observations, data reduction, and archival data sets}\label{sec:obs_and_dataredu}

 \subsection{Polarimetric observations of LDN 1225}\label{subsec:polarimetry}


Polarimetric observations were carried out
on five nights (2010 November 12 and 13, 2010 December 14, and 2013 November 01 and 02), using the ARIES Imaging Polarimeter \citep[AIMPOL;][]{Rautelaetal2004} 
mounted at the Cassegrain focus of the 104-cm Sampurnanand telescope of the Aryabhatta
Research Institute of observational sciencES (ARIES), Nainital, India.
The observations were carried out in the $R_{c}$ ($\lambda_{Rc_{eff}}$=0.670$\mu$m)
photometric band using a small area (370$\times$370 pixel$^2$)
of the TK 1024 $\times$ 1024 pixel$^2$ CCD camera. 
Field of view (FOV) of AIMPOL is $\sim$~8 arcmin diameter on the sky. During the observations the FWHMs of the stellar images  vary between 3~--~5$\arcsec$.
The read-out noise and gain of the CCD are
7.0 $e^{-1}$  and 11.98 $e^{-1}$/ADU respectively.
Since AIMPOL does not have the grid, we manually checked for any overlap of ordinary
and extraordinary images of the sources.
Fluxes of ordinary ($I_{o}$) and extra-ordinary ($I_{e}$) beams for all the observed sources
with good signal-to-noise ratio were extracted by standard aperture photometry
after bias subtraction using the {\sc IRAF}\footnote[1]{{\small IRAF} is distributed by National Optical
Astronomical Observatories, USA.} package. 
More details on the instrument and the detailed procedures used to 
estimate the polarization measurements are mentioned in \citet[][and references therein]{Eswaraiahetal2011}. 

All the measurements are corrected for both the instrumental polarization and offset polarization 
angle by observing unpolarized and polarized standard stars, respectively, 
from \citet{Schmidtetal1992}. 
As given in the Table \ref{tab:polstandresults}, our results on the polarized standard stars are in good agreement, 
within the observational errors, with those from \citet{Schmidtetal1992}. 
Measurements of one unpolarized standard star, HD\,21447, 
as listed in the Table \ref{tab:polstandresults}, show that the instrumental 
polarization in $R_{c}$-band is $\leq$~0.1$\%$. 
This is consistent with the fact that the instrumental polarization of AIMPOL 
has been monitored since 2004 on various observing nights as a part of various projects and found to be less than 0.1\% in different bands \citep{Rautelaetal2004,Medhietal2008,Eswaraiahetal2011,Eswaraiahetal2012,Pandeyetal2013,Eswaraiahetal2013,Kumaretal2014b,Kumaretal2016,Leeetal2018}. 

We have obtained the $R$-band polarizations (degree of polarization, $P_{R}$ (\%), and polarization angle, $\theta_{R}$ ($\degr$)) 
of 280 stars and are listed in Table \ref{tab:pol280stars}. 
This table also contains the stellar coordinates and 2MASS photometric data. 

Figure \ref{fig:polvecmap_colorcompo}(a) displays the polarization vector map 
towards LDN\,1225 using $P_{R}$ and $\theta_{R}$ of 280 stars, which depicts that B-field geometry is not uniform.
Of 280, 33 stars are distributed within the 8$\arcmin$~diameter field containing the star forming region NGC\,7538 which is situated within the Perseus spiral arm of our Galaxy at a distance of 2.65 kpc \citep{Moscadellietal2009,Pugaetal2010,Sharmaetal2017}. 
Since these stars are not physically associated with LDN\,1225 (located at $\sim$800~pc), we exclude them from further analyses. 
We also omitted 9 stars with NIR-excess ($[J-H] \geq 1.69 \times [H-K]$; see e.g., \citealt{Eswaraiahetal2013,Eswaraiahetal2017}) as their polarizations might be consisting of intrinsic components due to asymmetric distribution of material in their circumstellar disks. Remaining 238 stars are used in the further analyses. 
 
\subsection{Polarimetric observations of several fields covering parts of CepOB3}\label{sec:polobs_cepob3_fields}

To understand whether the large scale B-fields of CepOB3 are preserved and have any effect on the small scales of LDN\,1225, 
we also observed 19 additional fields covering different parts of CepOB3 using various polarimeters. 
Three fields with FOV of 8$\arcmin$ diameter were 
observed in $R$-band with AIMPOL on 2016 October 29 and the data are reduced by adopting similar procedures mentioned in the above section \ref{subsec:polarimetry}. Eight fields with FOV of 4$\arcmin$ diameter were observed in $R$-band on five nights (2014 November 19, 20, 25, 26, and 27) using the 2-m telescope of the Inter University Center for Astronomy and Astrophysics (IUCAA), Girawali Observatory, India. The instrument used was the IUCAA Faint Object Spectrograph and Camera (IFOSC) in the polarimetric mode \citep{SenTandon1994,Ramaprakashetal1998}. Eight fields with FOV of $4\arcmin\times4\arcmin$ are observed on 2015 October 13 with Triple Range Imager and Polarimeter (TRIPOL) mounted on Lulin One-meter Telescope (LOT) at Lulin observatory, Taiwan (Sato et al. in prep.). TRIPOL acquired simultaneous observations in SDSS $g^\prime$-, $r^\prime$-, and $i^\prime$-bands, but we use only the $r^\prime$-band data. 
The measurements from IMPOL and TRIPOL are corrected for both instrumental polarization and offset polarization angles. 

Apart from our own observations, we use catalogued $V$-band polarization \citep{Heiles2000} of 16 stars located within the $5\degr\times5\degr$ field containing CepOB3. We make sure that this sample does not include the stars with possible intrinsic polarization (e.g., Be stars, other emission type stars, young stellar objects, super-giant, red/blue supergiant, Wolf-rayet stars, etc.) with the help of SIMBAD database.

More details on the results and discussion based on the polarizations of CepOB3 are given in 
Sections \ref{sec:b_fields_cepob3_opt} and \ref{subsec:associationofldn1225withcepob3}.

 \subsection{Distances from GAIA DR2}\label{subsec:gaia_dist}

The ESA GAIA mission \citep{Gaiacollaboration2016a,Gaiacollaboration2016b} 
data release 2 \citep[][hereafter GAIA DR2]{Gaiacollaboration2018b}, provides distance information for stars up to 21 mag. 
 Of the 238 stars which have been used in the present study, 197 have GAIA DR2 parallaxes passing 10$\sigma$ criterion (i.e., 
 parallaxes to their uncertainties, $\omega/\sigma_\omega$, $\geq$~10). 
 The search area of 1$\arcsec$ radius around each of the star was used while matching the coordinates of our stars with those of GAIA DR2. Distances of all 197 stars are derived using the relation $d$~$=$~(1000/$\omega$) pc (where $\omega$ is the parallax in milli-arcsecond) and the distance error is estimated by propagating the distance-parallax relation. Based on the asteroseismic data of well-characterized red giant branch (RGB) stars in the {\it Kepler} field, 
 \citep[][and references therein]{Zinnetal2018} have independently confirmed the offset in the parallaxes, $\sim$52.8~$\mu$as (micro-arcsecond), of GAIA DR2. However, we have not applied the offset in this study as our main concern is to find the distance of the cloud rather than deriving the distances of individual stars. 
The distances of the stars are given in the column 9 of Table \ref{tab:pol280stars}. 

\subsection{Photometric observations towards LDN 1225}\label{subsec:photometry}
Optical photometric observations were performed on 2013 September 6 and 7 using Himalaya Faint Object Spectrograph and Camera (HFOSC) of the 2-m Himalayan Chandra Telescope (HCT) of Indian Astronomical Observatory (IAO), Hanle, operated by Indian Institute of Astrophysics (IIA), India. The central 2048$\times$2048 pixel$^{2}$ area of the 2K$\times$4K CCD was used for data acquisition.  The FOV of the instrument is $10\arcmin\times10\arcmin$ area with a pixel size of 15\,$\mu$m and image plate scale of 0.293$\arcsec$~pixel$^{-1}$. Photometric observations in $BVRI$-bands were performed towards four fields around LDN\,1225. We obtained multiple frames in each filters and the total integration time in $B$,$V$, $R$, and $I$-filters was 300, 200, 90, and 90 sec, respectively, in each field. Typical seeing during the observations was $\sim$1$\farcs$5. The total area coverage of our optical photometric observations are of $\sim$17$\arcmin\times20\arcmin$ around LDN\,1225, which is shown with a white box in Figure \ref{fig:polvecmap_colorcompo}(a). In order to apply  atmospheric extinction correction as well as for instrument calibration, we observed the  standard star field PG0231 \citep{Landolt1992} during the nights. PG0231 was observed at various  air masses  and the nights were photometric. Bias and flat frames were also acquired during the nights. 

After bias and flat correction, we stacked all the images of individual fields in each filter. We performed point spread function (PSF) photometry using the {\sc DAOPHOT} package in {\sc IRAF}  to derive the  instrumental magnitudes. Extinction and instrument coefficients were measured using the instrumental magnitudes and  standard star magnitudes of PG0231 
field \citep{Landolt1992}. These coefficients were applied to the instrumental magnitudes of LDN 1225 and thus we obtained the final, calibrated photometry. The individual catalogs of four fields were then merged to obtain the final catalog. Typical uncertainty in the photometric calibration is $\sim$ 0.05, 0.03, 0.02, and 0.02 mag for the $B$, $V$, $R$ and $I$ filters.  

Table \ref{tab:phot689stars} lists 689 stars with optical ($BVRI$) plus NIR ($JHK_{s}$; from 2MASS catalog) photometric data with uncertainties less than 0.1 mag. 
These data have been used in color-color diagrams to investigate the nature of extinction law in the foreground and cloud mediums (cf., Appendix~\ref{subsec:rv_estimation}).

 \subsection{$^{12}$CO (1~--~0),  $^{13}$CO (1~--~0), and C$^{18}$O (1~--~0) data from PMO MWISP survey}\label{pmo_co_data}
 
 In order to study the gas kinematics of the cloud LDN\,1225, we used the $^{12}$CO(1~--~0), $^{13}$CO (1~--~0), and C$^{18}$O (1~--~0) molecular line data from the Milky Way Imaging Scroll Painting (MWISP) project\footnote{\url{http://www.radioast.nsdc.cn/mwisp.php}} using the Purple Mountain Observatory (PMO) Delingha 13.7 m telescope located at Qinghai observing station \citep[see][]{Suetal2016}. LDN\,1225 has been observed in November 2015. The MWISP project is a new large-scale survey aiming to perform molecular line observations at the $J$~=~1~--~0 transition of CO isotopes $^{12}$CO, $^{13}$CO, and C$^{18}$O simultaneously. This survey has been specially intended to observe the northern Galactic plane within the Galactic coordinates of $-$10$\fdg$~25~$\leq$~$l$~$\leq$~250$\fdg$25 and $-$5$\fdg$25~$\leq$~$b$~$\leq$~+5$\fdg$25, and several other regions of interests. One of the main goals of the MWISP project is to investigate the physical properties of molecular clouds along the northern Galactic plane. 
The beam sizes are about $49\farcs8$ and $51\farcs3$ at 115.3 GHz and 110 GHz, respectively. 
The nominal sensitivities (rms level) in the brightness temperature are 0.47 K in $^{12}$CO at the velocity resolution of 0.159 km~s$^{-1}$ and 0.26 K in $^{13}$CO and C$^{18}$O at the velocity resolution of $\sim$0.167 km~s$^{-1}$. The resultant data cubes have a FOV of 30$\arcmin\times30\arcmin$ and a beam size of $\sim$53$\arcsec$. Our target LDN\,1225 is located within the MWISP survey area data of CepOB3.
More details on the instrument, data reduction, and analysis procedures can be found at \citet{Gongetal2016} and \citet{Suetal2017}.
 
\section{Analyses and Results}\label{sec:analyses_and_results}

\subsection{Distribution of polarization measurements}\label{subsec:distriofppa}

Figure \ref{fig:p_vs_pa} plots the $P_{R}$ versus $\theta_{R}$ for 238 stars. 
The $\theta_{R}$ values increase from $\sim$50$\degr$~to~$\sim$75$\degr$ as a function of increase in $P_{R}$ from $\sim$1\%~to~$\sim$2\%. Beyond which, 
$P_{R}$ increases up to $\sim$5\% while the $\theta_{R}$ values are distributed between $\sim$75$\degr$ and $\sim$110$\degr$. 
To probe the B-field orientation of LDN\,1225 we require background (hereafter BG) stars. 
For this, we need to identify and exclude the FG stars and stars that are not physically 
associated with the local arm in which LDN\,1225 is located. 
Though, there exist two clear distributions in $P_{R}$ versus $\theta_{R}$ diagram 
(separated at $P_{R}$~$=$~2\% and $\theta_{R}$~$=$~75$\degr$ shown with dashed lines in Figure \ref{fig:p_vs_pa}, 
also see Figure \ref{fig:polvecmap_colorcompo}(a)), 
identification of stars belonging to each distribution will be difficult. 
This can be overcome by knowing the stellar distances, cloud distance, and by classifying the entire sample into FG and BG stars based on their polarizations and distances.


\subsection{Distance versus polarization plots and distance estimation of LDN\,1225}\label{subsec:distanceldn1225}

Nearly 93\% (197 out of 238) of the stars with polarization data have distance information satisfying the criterion $\omega/\sigma_{\omega}\geq10$ (cf. section \ref{subsec:gaia_dist}). This information is highly useful to estimate the accurate distance of the cloud. The basis for the distance determination is the occurrence of an abrupt jump in the level of polarization at the cloud distance. The degree of polarization as a function of distance, in the presence of uniform distribution of material 
along the line-of-sight and according to the canonical extinction to distance relation of $\langle A_{V}/L\rangle$~$\approx$~1.8 mag kpc$^{-1}$ \citep{Whittet2003}\footnote{This relation is applicable only as a general average for lines of sight close to the plane of the Milky Way and for distances up to a few kiloparsecs from the Sun.}, should exhibit a monotonically increasing trend. 
However, the presence of a cloud (hence increase in the number density of aligned dust grains of the cloud) along the line-of-sight through which background stellar light passes by would cause an enhanced level of polarization at the cloud distance. As a result, a sudden jump in the level of polarization would prevail at the cloud distance.  
Any change in the orientation of B-fields can also be witnessed in the distance versus polarization angle plot if the B-fields of the cloud are different from those of the FG medium.

Figures \ref{fig:dist_vs_p_pa_q_u}(a) and (b), respectively, show the distance 
versus $P_{R}$ and $\theta_{R}$ of 197 stars. 
Both $P_{R}$ and $\theta_{R}$ seem to increase as a function of distance up to 830~pc. 
In contrast, beyond 830~pc a sudden jump followed by a scattered distribution in $P_{R}$ and 
a constant trend in $\theta_{R}$ can be witnessed. 
Similarly, Stokes parameters ($Q_{R}$ and $U_{R}$) also exhibit linearly decreasing or
increasing trends up to 830~pc and beyond which they exhibit scattered distributions 
(Figures \ref{fig:dist_vs_p_pa_q_u}(c) and (d)). These abrupt change in $P_{R}$, $\theta_{R}$, $Q_{R}$, and $U_{R}$ at 830~pc is attributed to the
presence of LDN\,1225 at this distance.

Since we have used stars with 10$\sigma$ GAIA DR2 distances in our analyses, the expected measurement uncertainty 
at 830~pc would be 83~pc (10\% of 830~pc). Therefore, we ascertain the distance of LDN\,1225 as 830$\pm$83~pc. 
This value supports the distance estimation of 730\,$\pm$\,120~pc by other studies 
\citep[][see also \citealt{Blaauwetal1959,CrawfordBarnes1970,Sargent1977,LaddHodapp1997,Hartiganetal2000,Kunetal2008}]{Bruntetal2003}.

In order to elucidate more on the above distance determination, 
we plot the differential polarization ($P_{i+1}-P_{i}$), polarization angle ($\theta_{i+1}-\theta_{i}$),
and Stokes ($Q_{i+1}-Q_{i}$ and $U_{i+1}-U_{i}$) versus
distance ($Distance_{i+1}$) parameters as shown in Figure \ref{fig:diffppaqu}. The subscript {\it i} varies from 1 to~196~in
the increasing order of distances. Since the number of stars is 197, the total differential measurements would be 196. 
This plot offers a crucial information on the changes in the patterns of polarization measurements of any star relative to those of an immediate foreground. 
However, in order to see the clear variations of the parameters at the cloud's distance, 
we limit the data samples up to 2 kpc and plotted distance in linear scale. 

Evidently all the differential parameters exhibit a clear transition from negative to positive (or vice versa) 
at 830~pc as denoted with a green arrow. Region I is dominated by the FG stars, while regions III and IV are with BG stars. 
Stars with 747 (830~$-$~83)~pc~$<$~$Distance_{i+1}$~$<$~830~pc (blue circles) are lying in the region I only, in contrast those 
with 830~pc~$<$~$Distance_{i+1}$~$<$~913 (830~$+$~83)~pc (red circles) are clearly distributed in both regions III and IV. This observed 
feature suggests an evidence for 
the clear changes in the polarization properties of the stars at 830~pc. These changes we attribute to the presence of cloud LDN\,1225 at 830~pc.

\subsection{Identification of foreground and background stars, and their polarization characteristics}\label{subsec:fgbgidentification}

Since the distance of LDN\,1225 is constrained as 830$\pm$83~pc (above section), we consider 61 stars 
with distances $<$ 830~pc as foreground stars (hereafter FG) and are mentioned as ``FG" in column 10 of 
Table \ref{tab:pol280stars}, whereas the 136 stars with distances $\geq$ 830~pc as background stars. 
However, there exist a star forming region NGC\,7538 
(situated within the Perseus arm at 2.65 kpc; \citealt{Moscadellietal2009,Pugaetal2010,Sharmaetal2017}) 
lying background to LDN\,1225 but projected close ($\sim$15$\arcmin$) to it. 
In addition, we note below that there exist a background cloud component at the same location of LDN\,1225, which 
is termed as {\it background cloud} of LDN\,1225. In order to examine the physical connection of this component 
with NGC\,7538, we have constructed the position-velocity (PV) plots along the two cuts (cf., Figure \ref{fig:polvecmap_colorcompo}(b)) 
using $^{12}$CO data as shown in Figures \ref{fig:pvcuts}(a) and (b). 
Evidently, the emissions from LDN\,1225 and its {\it background cloud} are concentrated, respectively, 
at $V_{LSR}$~$=$~$-$11.2 km s$^{-1}$ and $V_{LSR}$~$=$~$-$54.5 km s$^{-1}$. These two components are located 
at the same position of the sky (distributed between two horizontal dashed lines) but at different distances 
(correspond to two dashed vertical lines). Interestingly, the $^{12}$CO emission from NGC\,7538 has been 
widely distributed between $\sim-$40.6~km~s$^{-1}$ and $\sim-$60~km~s$^{-1}$ with a prominent peak 
emission at $\sim-$56~km~s$^{-1}$, which is close to the $-$54.5 km s$^{-1}$ component from {\it the background cloud}. 
Moreover, the existence of faint $^{12}$CO emission (green contours around $V_{LSR}$~$=$~$-$54.5 km s$^{-1}$ in 
Figures \ref{fig:pvcuts}(a) and (b)) between {\it background cloud} and NGC\,7538 suggests a physical connection between them. 
Therefore, hereafter we consider {\it background cloud} of LDN\,1225 is nothing but the region of NGC\,7538 lying at 2.65 kpc.

Based on the above paragraph, we consider 18 stars with distances $>$ 2.65 kpc as the background stars 
of NGC\,7538 and are mentioned as ``BG-NGC\,7538" in the column 10 of Table \ref{tab:pol280stars}. 
These stars are excluded from the further analysis, because their polarizations might have comprised of 
the dust properties and B-field orientations of both clouds NGC\,7538 and LDN\,1225. 
The remaining 118 stars having distances between 830~pc~--~2.65~kpc, as highlighted within the yellow 
background in Figure \ref{fig:dist_vs_p_pa_q_u}, were considered as the true background stars representing 
the dust properties and B-field orientation of the cloud LN\,1225. These stars are mentioned 
as ``BG" in column 10 of Table \ref{tab:pol280stars}.

Gaussian fits are performed to the polarization distributions of 61 FG and 118 BG stars as shown 
in the Figure \ref{fig:histogauss_p_pa}. Resultant mean and standard deviation of $P_{FG}$ and $\theta_{FG}$ 
for FG stars are 1.4\,$\pm$\,0.4\% and 61\,$\pm$\,11$\degr$ and of $P_{BG}$ and $\theta_{BG}$ 
for the BG stars are 2.8\,$\pm$\,1.0\% and 87\,$\pm$\,11$\degr$, respectively.

\subsection{Foreground interstellar polarization (ISP)}\label{subsec:ispestimation}

As shown in Figures \ref{fig:dist_vs_p_pa_q_u}(c) and (d), the Stokes parameters of 61 FG stars follow a monotonically 
increasing trend by following a straight line. We took the advantage of this clean set of data to evaluate the 
foreground interstellar polarization (ISP) component. We performed weighted linear fits (red thick lines) 
to the distance versus $Q_{R}$ and $U_{R}$ parameters. The resultant parameters at 830~pc are $Q_{ISP}$~$=$~$-$0.9\,$\pm$\,0.1\% and $U_{ISP}$~$=$~1.4$\pm$0.1\% and the corresponding $P_{ISP}$ and $\theta_{ISP}$ values are 1.7$\pm$0.1\% and 61$\pm$2$\degr$. These values are 
considered as the ISP contribution. 
Similar methodology has been adopted to estimate the ISP contribution and retrieve the intrinsic polarizations 
of nearby clouds \citep[eg.,][]{Eswaraiahetal2013,Nehaetal2018}, 
B[e] stars \citep[eg.,][]{Leeetal2018}, and Supernova (Kumar, B., et al. under review). 

\subsection{Foreground polarization correction to infer B-field structure in LDN\,1225}\label{subsec:fg_subtraction}

In order to retrieve the true polarization contributions from the cloud LDN\,1225, 
we subtract ISP Stokes parameters ($Q_{ISP}$~$=$~$-$0.9$\pm$0.1\%, $U_{ISP}$~$=$~1.4$\pm$0.1\%) vectorially from those ($Q_{BG}$ and $U_{BG}$) of 118 BG stars and obtained intrinsic Stokes parameters ($Q_{C}$ and $U_{C}$) of the cloud using the following relations. 
\begin{eqnarray}
Q_{C}~=~Q_{BG}~-~Q_{ISP} \\
U_{C}~=~U_{BG}~-~U_{ISP}
\end{eqnarray}
We then estimate 
\begin{eqnarray}
	P_{C}~=~ \sqrt{Q_{C}^{2}+U_{C}^2} \\
	\theta_{C}~=~0.5~\arctan({U_{C}/Q_{C}}). 
\end{eqnarray}
Here, the subscript `C' stands for the cloud, LDN\,1225, component. The $P_{C}$ values lie between 0.3\%~--~6.8\%, with a Gaussian mean and dispersion of 2.4$\pm$1.0\%. 
The $\theta_{C}$ values lie between 63~--~128$\degr$, with a Gaussian mean and dispersion of 106$\pm$11$\degr$. This small dispersion in polarization angles implies an uniform B-field orientations within the parts of the cloud. The Gaussian fits to the $P_{C}$ and $\theta_{C}$ are shown in Figures \ref{fig:histogauss_p_pa_of_bg_c_comp}(a) and (b). 

In order to test if ISP has any significant impact on the cloud polarization components, 
we have constructed the histograms of $\Delta(P)$~$=$~$P_{C}$~$-$~$P_{BG}$ and $\Delta(\theta)$~=~$\theta_{C}$~$-$~$\theta_{BG}$ as shown in Figures \ref{fig:histogauss_p_pa_of_bg_c_comp}(c) and (d). The Gaussian fit resultant mean and dispersion of $\Delta(P)$ are $-$0.5\% and 0.6\% and similarly the mean and dispersion of $\Delta(\theta)$ are 14$\degr$ and 4$\degr$, respectively. This implies that the FG medium has a depolarization effect by $-0.5\%$ on the cloud's polarization and a systematic rotation effect by $14\degr$ on the cloud's B-field component.



\begin{figure*}
\resizebox{8.75cm}{8.75cm}{\includegraphics{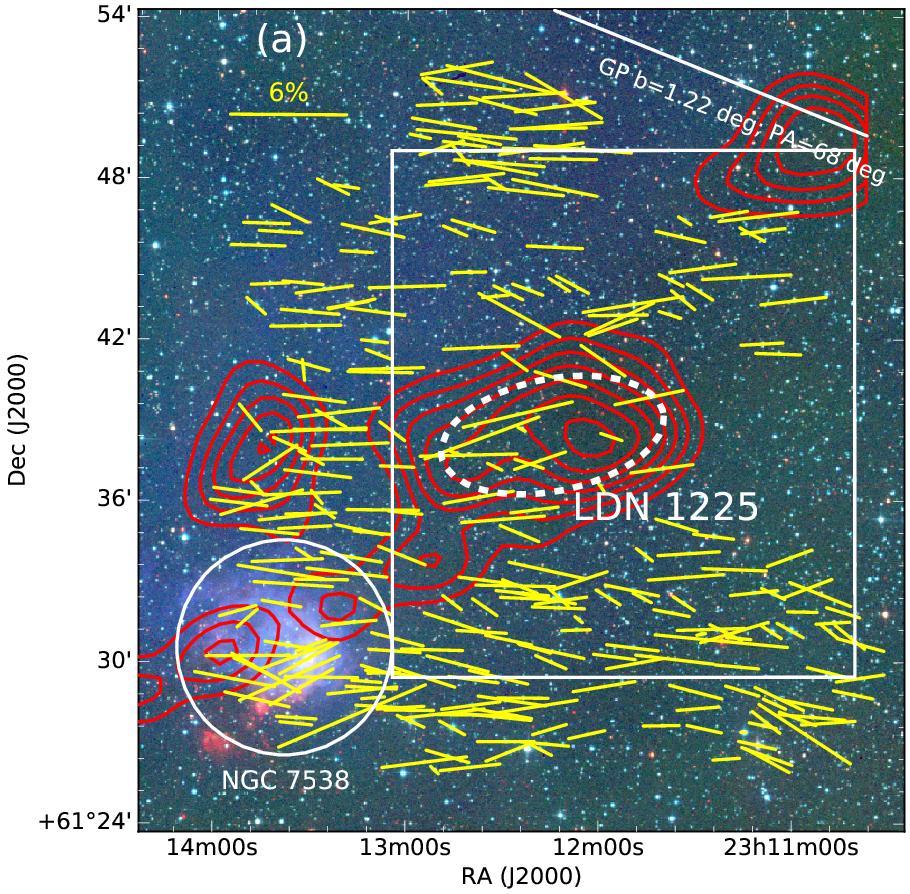}}
\resizebox{8.75cm}{8.75cm}{\includegraphics{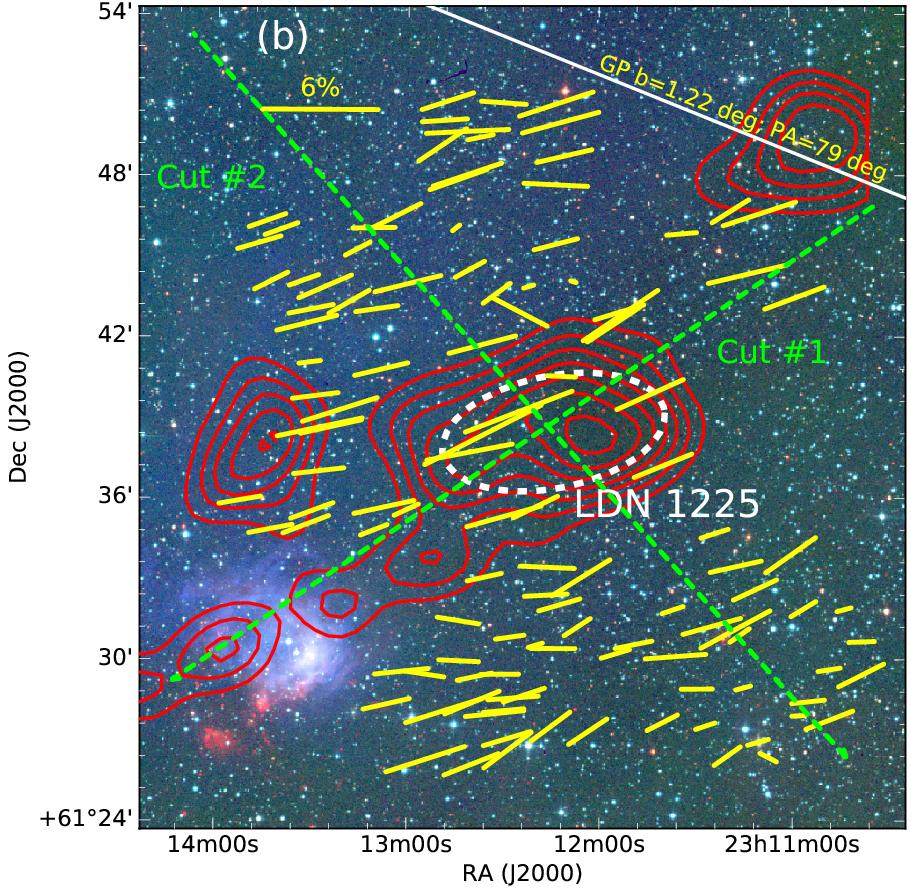}}
	\caption{B-field orientation towards LDN\,1225 before and after exclusion of unwanted stars (such as FG and BG-NGC\,7538 stars, stars around NGC\,7538, NIR excess sources) and correction of ISP from the polarizations of BG stars. (a) Polarization vector map of the region containing LDN\,1225, using $P_{R}$ and $\theta_{R}$ values of all 280 stars. Background is the color-composite of 2MASS $K_{s}$- (red), DSS I- (green) and DSS R-band (blue) images. Red contours represent the $^{12}$CO total integrated intensity map, drawn at 3, 6, 9, 12, 15, 18, 21, and 24 K km s$^{-1}$. Location of the dark globule LDN\,1225 and the star-forming region NGC\,7538 (with a circle of 8$\arcmin$ diameter) are shown and labeled. White dashed ellipse denotes the extent (major axis~$=$~8.4$\arcmin$ diameter, minor axis~$=$~4.2$\arcmin$ diameter, and position angle~$=$~102$\degr$) of LDN\,1225 based on the CO data (see Figure \ref{fig:moments012_ldn1225}). Reference vector with $P_{R}$~=~6$\%$ and $\theta_{R}$~=~90$\degr$ is shown. Galactic plane making a position angle (PA) of 68$\degr$ towards LDN 1225 is shown with a white line. White box with the dimensions of $\sim17\arcmin\times20\arcmin$ marks the region around LDN\,1225 selected for optical photometry in $BVRI$-bands. (b) Same as (a) but depicts the B-field map of LDN\,1225 cloud using the polarization measurements ($P_{C}$ and $\theta_{C}$) of 118 BG stars having accounted for interstellar polarization (see section \ref{subsec:fg_subtraction}). Two PV cuts are shown and labeled. Cut 1 is parallel (with position angle of $\sim$124$\degr$) and cut 2 is perpendicular (with position angle $\sim$222$\degr$) to the elongation of the LDN\,1225.} \label{fig:polvecmap_colorcompo}
\end{figure*}

\begin{figure*}
	\centering
\resizebox{8.5cm}{8.5cm}{\includegraphics{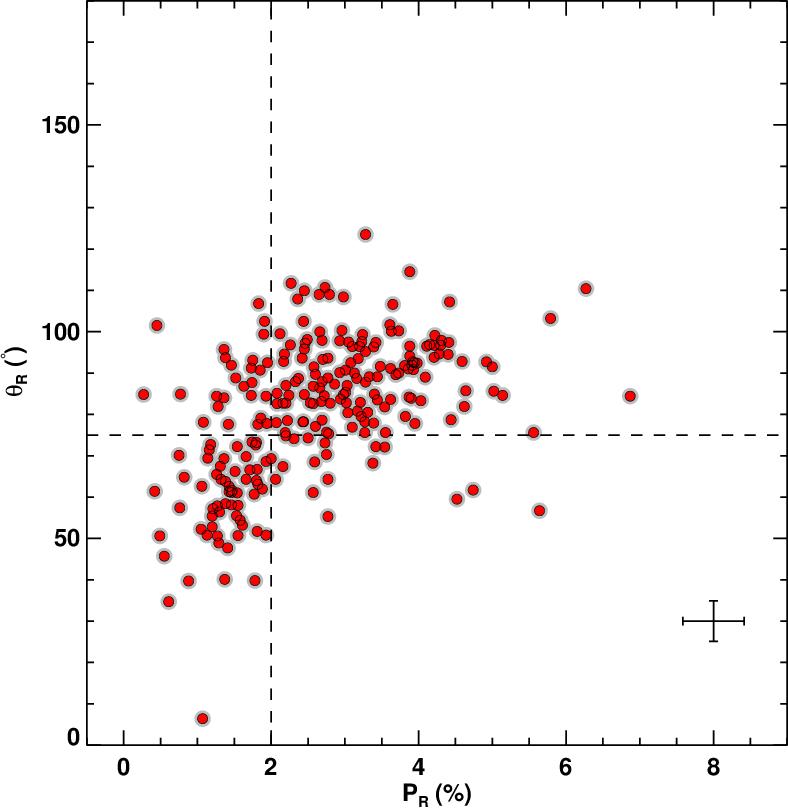}}
\caption{Polarization ($P_{R}$) versus polarization angle ($\theta_{R}$) of 238 stars located towards LDN 1225 using filled red circles. Two populations of stars with different values of $P_{R}$ and $\theta_{R}$ are evident and are separated by dashed lines drawn at $P_{R}$~=~2$\%$ and $\theta_{R}$~=~75$\degr$. Typical error is shown at the bottom-right side of the plot. \label{fig:p_vs_pa}}
\end{figure*}

%
Figure \ref{fig:polvecmap_colorcompo}(b) displays the B-field morphology of the LDN\,1225 region based on the polarization measurements ($P_{C}$ and $\theta_{C}$) of 118 BG stars as depicted with yellow vectors. 
It should be noted here that these measurements are corrected for ISP contribution. 
Based on the $^{12}$CO data, the position angle of the major axis of LDN\,1225 is 
found to be $\sim$102$\degr$ (depicted with a white dashed ellipse in Figure \ref{fig:polvecmap_colorcompo}; 
see Section \ref{districolines} and Table \ref{tab:cloudpas}) and the mean B-field orientation inferred 
by optical polarimetry ($\theta_{C}$) of the globule is 106$\pm$11$\degr$. 
This implies that the cloud structure (major axis of the globule) is nearly aligned parallel to the B-fields. 
This morphological correlation between the cloud structure and B-fields 
manifests an important role of B-fields in the formation of the globule, detailed discussion on which will be given in Section \ref{districolines}.

The B-field orientation of LDN\,1225, inferred from mean $\theta_{c}$\,$=$\,106\,$\pm$\,11$\degr$, is offset by 38$\degr$ from the position angle, $68\degr$, of the Galactic plane (GP) at $b$\,$=$\,1.22$\degr$ (white line in Figures \ref{fig:polvecmap_colorcompo}(a) and (b) and dashed vertical line in Figure \ref{fig:histogauss_p_pa}(b)). Nonetheless, within the error, the foreground component (mean $\theta_{FG}$~$=$~60\,$\pm$\,11$\degr$) is parallel to the GP component suggesting that the dust grains of the FG medium are aligned by the local B-fields that are parallel to the GP.


\subsection{Polarizations of FG and cloud mediums of CepOB3 based on optical polarimetry}\label{sec:b_fields_cepob3_opt}

By adopting the same distance criteria as of LDN\,1225, 
we identify FG and BG stars in 19 fields of CepOB3 (Section \ref{sec:polobs_cepob3_fields}). 
Furthermore, assuming that LDN\,1225 and CepOB3 share a common ISP contribution 
(cf. Section \ref{subsec:fg_subtraction}), we correct for it from the polarizations of 
BG stars and then derive the 
cloud polarizations of 19 fields of CepOB3. 
The resultant weighted mean polarizations of foreground and cloud mediums of these 19 fields 
and their central coordinates, number of stars in each field, 
and the instrument (or) source used are given in Table \ref{tab:weightedmeanPPAcepob3}. 

Similarly, $V$-band polarizations of 15 stars, distributed within the $5\degr\times5\degr$ 
area of CepOB3, are extracted from \citep{Heiles2000} catalog. Of these, 9 are FG and 6 are BG 
stars according to their distances from GAIA DR2 as well as our analyses concerning 
the identification of FG and BG stars (eg., Section \ref{subsec:fgbgidentification}). 
Their coordinates, HD numbers, and polarizations are given in Table \ref{tab:weightedmeanPPAcepob3}. 

A detailed discussion on the B-field morphologies of FG and cloud mediums of CepOB3 are given in 
Section \ref{subsec:associationofldn1225withcepob3}.

\subsection{Gas velocity dispersion and number density in LDN 1225 using PMO CO data}\label{subsec:gas_disp_numb_dense}

In order to evaluate the relative importance of B-field to turbulence and gravity, we need to estimate the B-field strength and its pressure. For this, we extract gas velocity dispersion ($\delta_{V_{LSR}}$) and number density ($n(H_{2})$) from the $^{12}$CO(1~--~0) molecular line data from the PMO MWISP project.
Figure \ref{fig:moments012_ldn1225} show the $^{12}$CO(1~--~0) total integrated intensity, velocity, and velocity dispersion maps. These maps were constructed using the velocity channels ranging from $\sim-$13 km s$^{-1}$ to $\sim-$3 km s$^{-1}$, and having the brightness temperature ($T_{b}$) above 3 times of the rms noise. This selection criterion on the range of velocity channels excludes CO emission from the background cloud NGC\,7538 (at $\sim-$54.5 km s$^{-1}$; cf., Figures \ref{fig:pvcuts}(a) and (b)). 

Using CASA software, a two-dimensional Gaussian is fitted to the CO velocity integrated intensity map of LDN\,1225 
and the resultant 
spatial extents of major and minor axes, and the position angle of the major axis are 
8$\farcm$6\,$\pm$\,0$\farcm$2, 4$\farcm$3\,$\pm$\,0$\farcm$1, and 102$\degr$\,$\pm$\,1$\degr$, respectively, 
and are represented with a black dashed ellipse in Figure \ref{fig:moments012_ldn1225}. These values 
are consistent with those (8$\farcm$4, 4$\farcm$2, and 102$\degr$) derived from FCRAO CO data as given in the vizier catalog\footnote{Available at \url{http://cdsarc.u-strasbg.fr/viz-bin/Cat?J/ApJS/144/47}}.

The mean integrated intensity (W(CO)) distributed within the extent of LDN\,1225 
(black ellipse, Figure \ref{fig:moments012_ldn1225}) is 17\,$\pm$\,4 K km s$^{-1}$. 
Using the CO-to-H$_{2}$ conversion factor, 
$X$~$\equiv$~$N(H_{2})/W(CO)$
~$=$~(2.0\,$\pm$\,0.6)$\times$10$^{20}$ 
cm$^{-2}$~(K~km~s$^{-1}$)$^{-1}$ 
\citep[][see also \citealt{Dameetal2001} 
and \citealt{SofueKataoka2016}]{Bolattoetal2013}, 
the mean column density $N(H_{2})$ is estimated as (3.4\,$\pm$\,1.3)$\times$10$^{21}$~cm$^{-2}$. 
Assuming the diameter of minor axis, 4$\farcm$3$\pm$0$\farcm$1 (equivalent to 1.04\,$\pm$\,0.02~pc at 830~pc), 
as the thickness of LDN\,1225, the mean number density, $n(H_{2})$, is (1.1\,$\pm$\,0.4)$\times$10$^{3}$ cm$^{-3}$.

Gaussian fit to the $^{12}$CO mean $T_{b}$ versus $V_{LSR}$ spectrum over an entire region of LDN\,1225 resulted a gas velocity dispersion $\sigma_{V_{LSR}}$~$=$~0.88\,$\pm$\,0.01 km s$^{-1}$. In addition, the fitted values of $V_{LSR}$ and $T_{b}$ are $-$11.08\,$\pm$\,0.01 km s$^{-1}$ and 4.96\,$\pm$\,0.04 K, respectively.
 
\subsection{B-field strength in LDN 1225}\label{sec:b_field_strength} 
Using the dispersion in the B-fields $\sigma_{\theta_{C}}$~$=$~11.4\,$\pm$\,0.3$\degr$ 
(cf. Section \ref{subsec:fg_subtraction}), 
gas velocity dispersion $\sigma_{V_{LSR}}$~$=$~0.88\,$\pm$\,0.01 km s$^{-1}$ 
(cf. Section \ref{subsec:gas_disp_numb_dense}), and the mean number density $n(H_{2})$~$=$~(1.1\,$\pm$\,0.4)$\times$10$^{3}$~cm$^{-3}$ 
(cf. Section \ref{subsec:gas_disp_numb_dense}) for LDN\,1225, we estimate B-field strength using the 
Davis-Chandrasekhar-Fermi method 
\citep[more commonly referred to as the Chandrasekhar-Fermi (CF) method;][]{Davis1951,ChandrasekharFermi1953}: 
\begin{equation}\label{cfrel}
B = Q~\sqrt{4\pi\rho}~\left(\frac{\sigma_{V_{LSR}}}{\sigma_{\theta_{C}}}\right). 
\end{equation}
The mass density $\rho$~$=$~$n(H_{2})$~$m_{H}$~$\mu_{H_{2}}$, where $n(H_{2})$ is the hydrogen volume density, $m_{H}$ is the mass of the hydrogen atom, and $\mu_{H_{2}}~\approx 2.8$ is the mean molecular weight per hydrogen molecule  which includes the contribution from helium. The correction factor $Q$~$=$~0.5 is included based on the studies using synthetic polarization maps generated from numerically simulated clouds \citep{Ostrikeretal2001} which suggest that for $\sigma_{\theta}$~$\leq$~25$\degr$, B-field strength is uncertain by a factor of two.
The estimated B-field strength is 56\,$\pm$\,10\,$\mu$G. Implications based on the relative importance of B-field to turbulent pressure in the formation and evolution of the globule LDN\,1225 is discussed in Section \ref{subsec:cloud_stability}.

\section{Discussion}\label{sec:discuss}

\subsection{Association of LDN\,1225 with CepOB3}\label{subsec:associationofldn1225withcepob3}
 
Based on a detailed comparisons of $V_{LSR}$, distance, and B-fields of LDN\,1225 with those of CepOB3, we discuss whether LDN\,1225 is associated with CepOB3. According to the $^{12}$CO and $^{13}$CO data (from PMO MWISP) of LDN\,1225, the mean $V_{LSR}$ values are $-$11.08\,$\pm$\,0.01 and $-$11.12\,$\pm$\,0.02 km s$^{-1}$, respectively. 
These values are consistent with the mean $V_{LSR}$ of $-$11.2 km s$^{-1}$ of a number of dark globules, namely, LDN\,1210, LDN\,1216, LDN\,1218, LDN\,1220, LDN\,1227, LDN\,1232, etc \citep{Yonekuraetal1997}, which are part of CepOB3. Similarly, several dense clumps of CepOB3 such as LDN\,1211, Cep A, Cep B, Cep C, Cep E, and Cep F are also found to have $V_{LSR}$ values with in a range of $-$8 to $-$12 km s$^{-1}$ \citep{Sargent1977,Yuetal1996}. These inferences suggest that LDN\,1225 is kinematically associated with the CepOB3. 

As per our present work, LDN\,1225 is located at the distance of 830$\pm$83~pc (c.f., section \ref{subsec:distanceldn1225}), which is consistent with the distance 700~pc assigned to LDN\,1225 assuming its association with CepOB3 \citep{Sargent1977,Dameetal1987,LaunhardtHenning1997}. The calibrated distance of CepOB3 via principal component analysis (under the assumption of an universal relationship between velocity dispersion and spatial scale within the clouds) on the 
Five College Radio Astronomy Observatory (FCRAO) Outer Galaxy Survey (OGS) CO data is 730\,$\pm$\,120~pc \citep{Bruntetal2003}. Using the methanol maser parallaxes, the distance of Cep A is ascertained to be 700\,$\pm$\,40~pc \citep{Moscadellietal2009}. This distance value is consistent with those quoted in other studies \citep{Blaauwetal1959,CrawfordBarnes1970,Sargent1977,LaddHodapp1997,Hartiganetal2000,Kunetal2008}. Therefore, LDN\,1225, located at 830$\pm$83~pc, is spatially and kinematically associated with the CepOB3.

Furthermore, using the polarization observations of 19 fields (Figure \ref{fig:obs_fields_fg_bg_bfields_cepob3}(a)), 
we compare the B-fields in the FG and cloud mediums of CepOB3 with those of LDN\,1225 to test 
if the large scales B-fields are preserved at smaller scales. 
B-field orientations of FG and cloud mediums of CepOB3, using our observations as well as 
\citet{Heiles2000} data (cf., Section \ref{sec:b_fields_cepob3_opt}), 
are shown in Figures \ref{fig:obs_fields_fg_bg_bfields_cepob3}(b) and (c). 
The weighted mean B-field orientation for the FG medium of CepOB3 is 67$\degr$ 
with a standard deviation of 26$\degr$ (Figure \ref{fig:obs_fields_fg_bg_bfields_cepob3}(b)), 
while for LDN\,1225 derived as 61$\degr$ with a standard deviation of 
11$\degr$ (cf., Section \ref{subsec:fgbgidentification}).
This implies that the B-fields in the FG mediums of both LDN\,1225 and CepOB3 are nearly similar, 
and are dominated by the GP component of 68$\degr$. 
Similarly, the weighted mean B-field orientation in CepOB3 is 113$\degr$ with a 
standard deviation of 29$\degr$ (as shown in Figure \ref{fig:obs_fields_fg_bg_bfields_cepob3}(c)), 
while that for LDN\,1225 is $106\degr$ with 
a standard deviation of 11$\degr$ (cf. Section \ref{subsec:fg_subtraction}). 
This result again suggests that, within the uncertainty, the mean B-fields in CepOB3 
and LDN\,1225 are similar implying that B-fields at the small scales of LDN\,1225 
are inherited from those at large scales of CepOB3. Hence, these large scale B-fields of 
CepOB3 could be important in governing the formation and evolution of LDN\,1225.

Therefore, in conclusion, LDN\,1225 is spatially, kinematically, and magnetically has similar characteristics as 
those for CepOB3, thereby confirming its association with CepOB3.


\subsection{Cloud stability}\label{subsec:cloud_stability}

To understand the importance of B-fields with respect to turbulence, we estimate the magnetic pressure and turbulent pressure using the relations $P_{B}$~$=$~$B^{2}/8\pi$ and $P_{turb}$~$=$~$\rho{\sigma_{turb}}^{2}$ (where $\sigma_{turb}$~$=$~$\sigma_{V_{LSR}}$) as (12\,$\pm$\,5)$\times$10$^{-11}$ dyn~cm$^{-2}$ and (4\,$\pm$\,1)$\times$10$^{-11}$ dyn~cm$^{-2}$, respectively. The mean $P_{B}$/$P_{turb}$ is estimated to be 3\,$\pm$\,2, suggesting that dominant role of B-fields over turbulence. 

To infer the importance of B-fields with respect to gravity, we estimate 
the mass-to-magnetic flux ratio in units of critical value using the 
following relation \citep{Crutcheretal2004,Chapmanetal2011}, 
\begin{equation}\label{eq:mtf}
\mu = \frac{(M/\phi)}{(M/\phi)_{crit}} =  7.6~N_{\parallel}(H_{2})/B_{tot},
\end{equation} 
where N$_{\parallel}$(H$_{2}$) is the column density, in units of 10$^{21}$ cm$^{2}$, along the magnetic flux tube and $B_{tot}$ is the total B-field strength in $\mu$G. Critical mass-to-flux ratio, ($M/\phi$)$_{crit}$~$=$~1/$\sqrt(4\pi^{2}G)$ \citep{NakanoNakamura1978}, corresponds to the stability criterion for an isothermal gaseous layer threaded by the perpendicular B-fields. The cloud region with $(M/\phi)$~$>$~$(M/\phi_{crit})$ i.e., $\mu$~$>$~1 will collapse under its own gravity and cloud is considered to be supercritical. The cloud with $\mu$~$<$~1 will be in a subcritical state because of the major support rendered by B-fields. Because of the projection effects between $N_{\parallel}$(H$_{2}$)/$B_{tot}$ and the actual measured $N$(H$_{2}$)/$B_{\parallel}$ (where $B_{\parallel}$ is the plane-of-the-sky B-field strength), B-fields parallel to the cloud structure, and assuming random orientation of B-fields with respect to the line of sight, the actual value of $\mu$ becomes (3/4)$\mu_{obs}$ 
\citep[][see their Appendix D.4\footnote{In order to correct for the projection effects on the estimated mass-to-flux ratio (in critical units), a factor 1/2 is valid for a spheroid cloud, 1/3 for an oblate spheroid flattened 
perpendicular to the B-fields, and 3/4 for a prolate spheroid elongated along the B-fields (similar to our case).}]{PlanckCollaborationetal2016}. 
Therefore, equation (\ref{eq:mtf}) can be re-written as $\mu$~$=$~5.7~$N(H_{2})/B_{\parallel}$. Using the $N$(H$_{2}$)~$=$~(3.4\,$\pm$\,1.3)$\times$10$^{21}$ cm$^{-2}$ (cf. Section \ref{subsec:gas_disp_numb_dense}) and $B_{\parallel}$~$=$~56\,$\pm$\,10\,$\mu$G (cf. Section \ref{sec:b_field_strength}) for LDN\,1225, $\mu$ is estimated to be 
0.35\,$\pm$\,0.15. This value suggests that LDN\,1225 is magnetically subcritical implying strong support by the B-fields, at least, for the outer low-density parts where optical polarimetry is reliable. However, the situation at relatively high dense regions may be different because of the dominance of gravity. 

  
Based on the NH$_{3}$ observations of LDN\,1225 by \citet*{ScappiniCodella1996}, 
the peak emission occurs at $V_{LSR}$~$=$~$-$11.5\,$\pm$\,0.1~km~s$^{-1}$, with a line width 
$\Delta_{V}$~$=$~0.8\,$\pm$\,0.1~km~s$^{-1}$. They consider LDN\,1225 as an inactive globule. 
However, as per the IRAS Serendipitous Survey Catalog \citep{Kleinmannetal1986}, within the 8$\arcmin$ diameter, LDN\,1225 hosts 
IRAS source IRAS 23094+6122 \citep[see also][]{MaheswarBhatt2006}. 
Additionally, in this study, based on mid-infrared (MIR) and far-infrared (FIR) images, 
we find that LDN\,1225 hosts two bright MIR YSOs and two {\it Herschel}/PACS 70\,$\mu$m sources (cf., Section \ref{subsec:70micronsources}). 
These sources are located in the dense parts of LDN\,1225, suggesting LDN\,1225 is not a quiescent cloud but on the verge of collapse and forming stars. 
  
 \subsection{Structure function analysis and the ratio of turbulent-to-order B-field strength}\label{subsec:structure_functionanalysis}


Structure function analysis (or angular dispersion function) has been used to derive the ratio of turbulent-to-order B-field strength (${\langle {B^{2}}_{t}\rangle}^{1/2}/B_{o}$). To separate the turbulent components from those of non-turbulence, we plotted $\langle\Delta{\theta^{2}}(l)\rangle^\frac{1}{2}$ \citep{Falceta-Goncalvesetal2008,Francoetal2010,Poidevinetal2010,Santosetal2012,Eswaraiahetal2013,FrancoAlves2015,Santosetal2016}, the square root of the second-order structure function or angular dispersion function (ADF)\footnote{The angular dispersion function (ADF) is defined as the square root to the average of the squared difference between the polarization
 angles measured for all pairs of points ($N(l)$) separated by a distance $l$.}, as a function of distance ($l$) as shown in Figure \ref{fig:structure_function}. We used the polarization angles ($\theta_{C}$) of 118 BG stars to compute
$\langle\Delta{\theta^{2}}(l)\rangle^\frac{1}{2}$, which depicts 
how the dispersion of the polarization
angles changes as a function of the length scale in LDN\,1225.

The square of the dispersion function can be approximated as follows \citep{Hildebrandetal2009}:
\begin{equation}\label{ADF_function}
\langle\Delta{\theta^{2}}(l)\rangle_{tot}-{\sigma^{2}_{M}}(l) = b^{2} + m^{2}l^{2}, 
\end{equation}
where $\langle\Delta{\theta^{2}}(l)\rangle_{tot}$ is the dispersion function computed from the data. The
quantity ${\sigma_{M}}^{2}(l)$ is the measurement uncertainties, which is simply the average of the variances 
on $\Delta{\theta(l)}$ in each bin. The quantity $b^2$ is the intercept of a straight line fit to the data
(after subtracting ${\sigma_{M}}^{2}(l)$). \citet{Hildebrandetal2009} have derived the equation for $b^2$
to find the ratio of turbulent to the large-scale magnetic field strength:

\begin{equation}\label{turb_large_b}
\frac{\langle B^{2}_{t}\rangle^{1/2}}{B_{o}}=\frac{b}{\sqrt{2-b^2}},
\end{equation}

In Figure \ref{fig:structure_function}, we show the ADF versus displacement using $\theta_{C}$ of 118 BG stars for LDN\,1225 region. The errors in each bin are similar to the size of the symbols. Each bin denotes $\sqrt{\langle\Delta{\theta^{2}}(l)\rangle_{tot}-{\sigma_{M}}^{2}(l)}$ i.e., the ADF corrected for the measurement uncertainties. Bin widths are in logarithmic scale. Only the first five data points were used in the linear fit to ensure that the length scale ($l$) in the fit (0.003 to 0.5~pc) is greater than the turbulent length scale ($\delta$) (which is of the order of 1 milli-parsec (mpc) or 0.001~pc; eg., \citealt{LiHoude2008}) and much shorter than the cloud length scale ($d \sim 1$ pc) i.e., $\delta~<~l~\ll~d$, to the data (equation (\ref{ADF_function})) versus distance squared.
The fitted function
is denoted with a thick dotted line.
Since our optical polarimetric observations have a low resolution due to the
available limited number of point sources,
the minimum length we probed is $\simeq$ 31 mpc.
The turbulent contribution to the total ADF is determined by the zero intercept of the fit to the
data at $l$\,$=$\,0. The net turbulent component, $b$, is estimated as 10\,$\pm$\,2$\degr$ 
(or 0.17\,$\pm$\,0.04 rad). This dispersion value (10\,$\pm$\,2$\degr$) estimated using structure function analysis is nearly similar to the dispersion in $\theta_{C}$ obtained using Gaussian fit ($\sigma_{\theta_{C}}$~=~11$\degr$; Section \ref{subsec:fg_subtraction}). The ratio of the turbulent to large-scale magnetic field strength ($\sigma(\theta) = \langle B^{2}_{t}\rangle^{1/2}/{B_{o}}$) is computed using equation (\ref{turb_large_b}) as 0.12\,$\pm$\,0.03. This suggests that the turbulent component of the field is very small compared with the non-turbulent ordered B-field component, i.e., $B_{t} \ll B_{o}$. This suggests that large scale ordered B-fields are dominant over the turbulent component in the LDN\,1225. 


\subsection{Correlations among the observed multiple components in CO gas and B-fields}\label{subsec:correlations_cogas_bfields}

\begin{figure*}
        \centering
\resizebox{8.5cm}{11cm}{\includegraphics{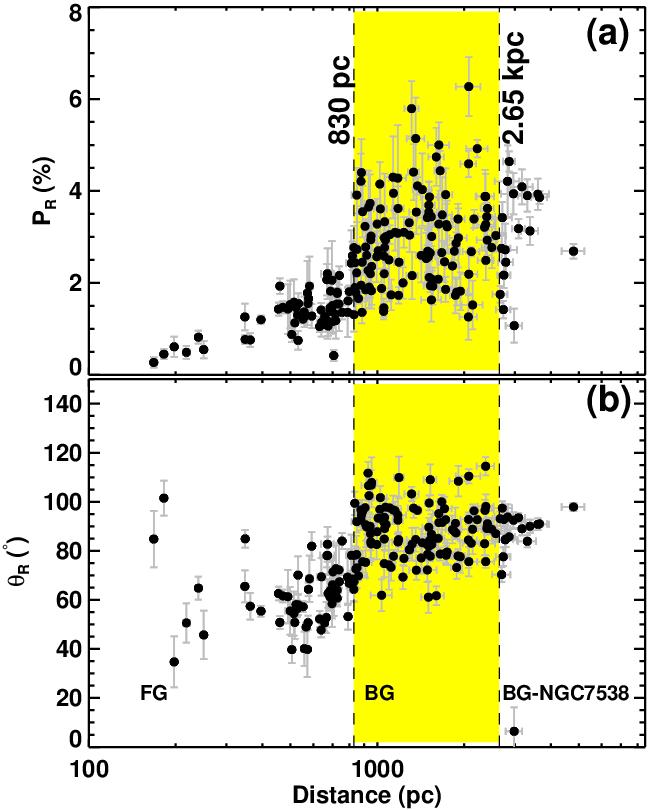}}
\resizebox{8.5cm}{11cm}{\includegraphics{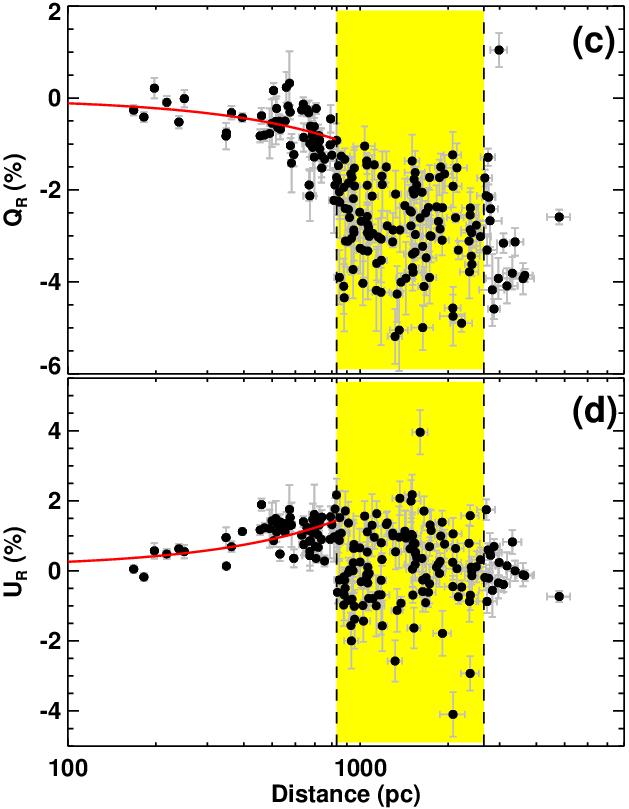}}
        \caption{Distance versus $P_{R}$ (a) and $\theta_{R}$ (b) of 197 stars satisfying the criterion 
	$\omega/\sigma_{\omega}\geq10$. Distance versus $Q_{R}$ (c) and $U_{R}$ (d) of the same stars 
	plotted in panels (a) and (b). Abrupt increase in $P_{R}$ and $\theta_{R}$ values, 
	and abrupt change in $Q_{R}$, and $U_{R}$ values are witnessed at a distance of 830~pc. 
	Weighted linear fits made to the $Q_{R}$ and $U_{R}$ versus distance values, 
	using 61 stars with distances $<$830~pc, are shown with red thick lines (see text for more details).
\label{fig:dist_vs_p_pa_q_u}}
\end{figure*}

\begin{figure*}
        \centering
\resizebox{16cm}{11cm}{\includegraphics{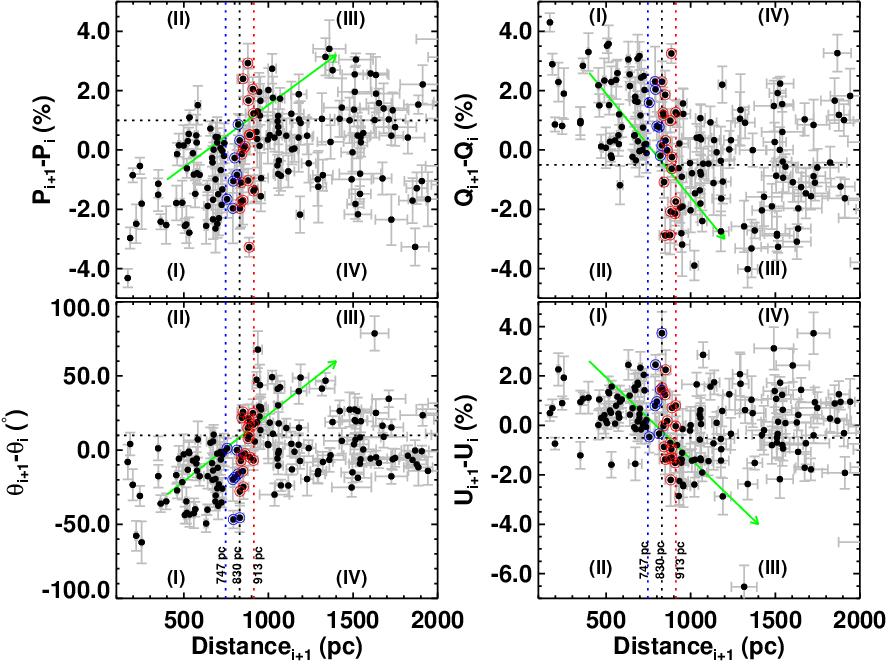}}
	\caption{Differential polarization ($P_{i+1}-P_{i}$), polarization angles ($\theta_{i+1}-\theta_{i}$), and 
	Stokes parameters ($Q_{i+1}-Q_{i}$ and $U_{i+1}-U_{i}$) versus distance ($Distance_{i+1}$) parameters for the stars used in 
	Figure \ref{fig:dist_vs_p_pa_q_u}. We used the data up to 2~kpc and distance in linear scale. 
	Black vertical (at 830~pc) and horizontal dotted lines in each panel separates the entire data samples into four regions. 
	Region II is devoid of the data (at most four stars in panel of $U_{i+1}-U_{i}$ vs $Distance_{i+1}$) 
	in comparison to those in other three regions (I, III, and IV). 
	In all the panels, region I is dominated by the FG stars, while the regions III and IV with BG stars.
	Blue and red dotted vertical lines are drawn, respectively, at 747~pc and 913~pc corresponding to the lower and upper limits of the ascertained 
	distance 830~pc with an uncertainity of 83~pc for LDN\,1225. Blue circles represent the stars distributed between 747~pc~$<$~$Distance_{i+1}$~$<$~830~pc, 
	while the red circles with 830~pc~$<$~$Distance_{i+1}$~$<$~913~pc. 
	Green arrow marks a clear transition of the differential parameters from negative to positive (or vice versa) at the distance of 830~pc. 
	\label{fig:diffppaqu}}
\end{figure*}

\begin{figure*}
        \centering
\resizebox{17cm}{10cm}{\includegraphics{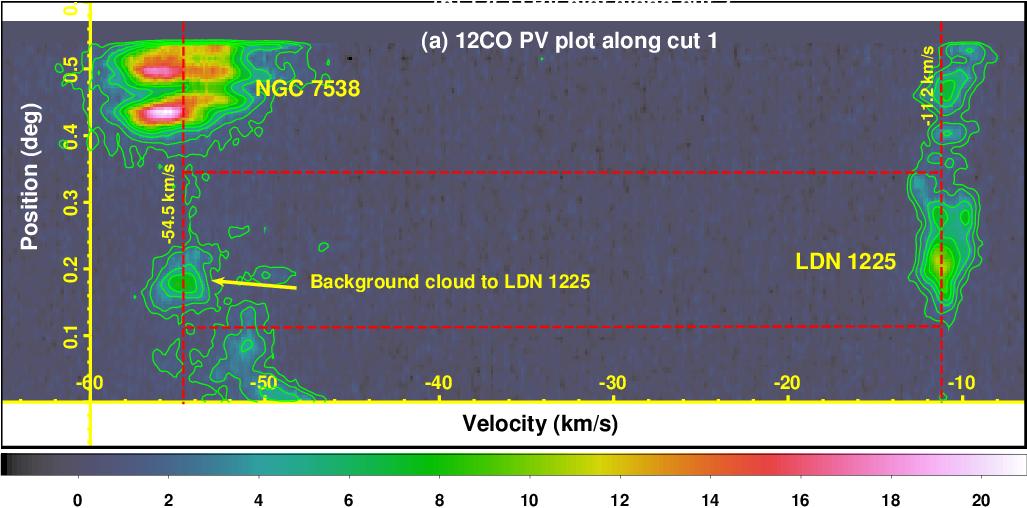}}
\resizebox{17cm}{10cm}{\includegraphics{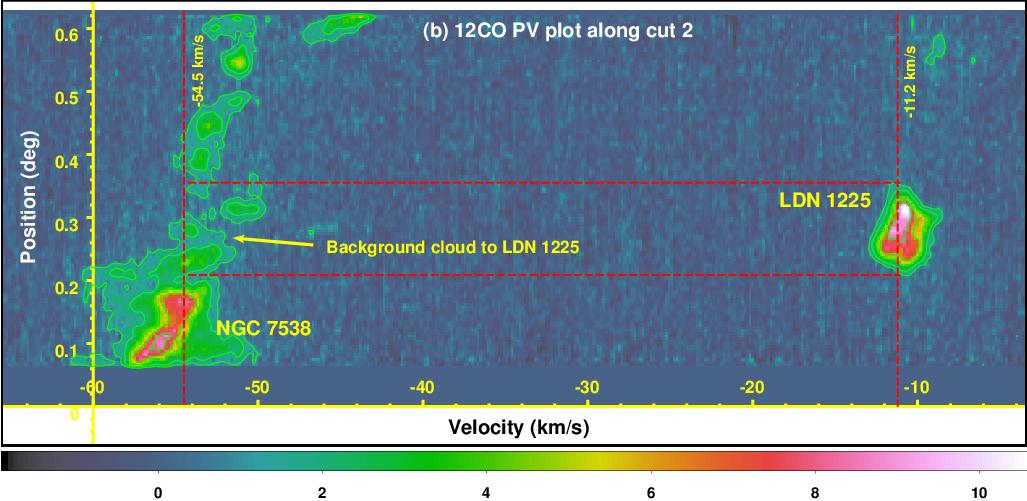}}
	\caption{$^{12}$CO position-velocity (PV) plots along the two cuts 
	shown in Figure \ref{fig:polvecmap_colorcompo}(b). 
	(a) PV plot for cut 1 (parallel to the elongation of LDN\,1225). 
	(b) Same as (b) but along the cut 2 (perpendicular to the elongation of LDN\,1225).
	Green contours, correspond to the $^{12}$CO intensity, are drawn at 1, 2, 4, 6, and 8 K. Vertical dashed lines
        drawn at $-$11.2~km~s$^{-1}$ and $-$54.5.2~km~s$^{-1}$, respectively, corresponds to LDN\,1225 and its background cloud located at similar
        velocity of NGC\,7538. Horizontal dashed lines depict the spatial extent of LDN\,1225 emission for a given cut. 
	\label{fig:pvcuts}}

\end{figure*}

\begin{figure*}
        \centering
\resizebox{8.5cm}{11cm}{\includegraphics{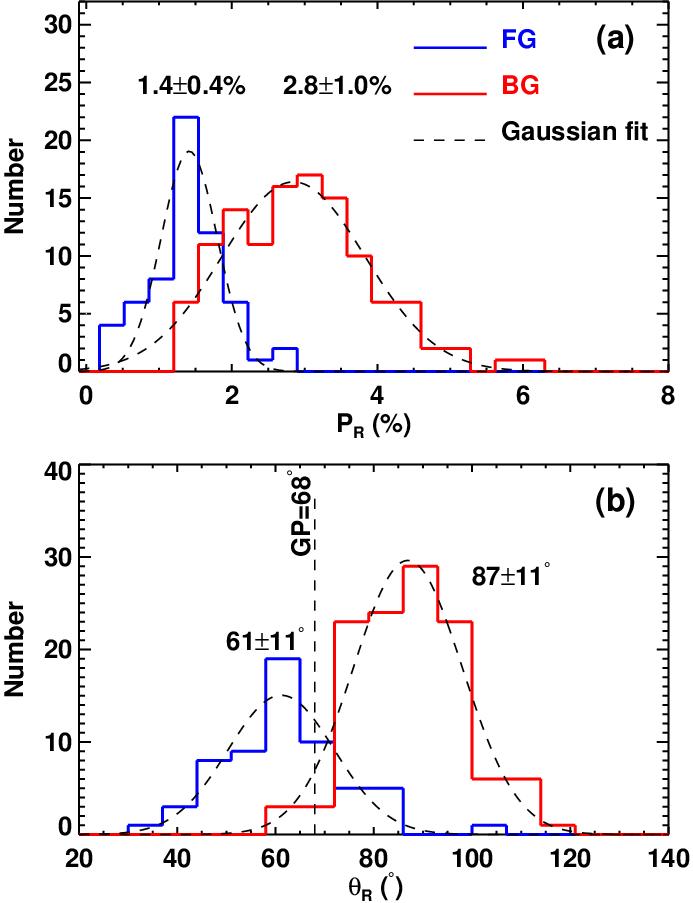}}
        \caption{(a) Number distribution of $P_{R}$ of the FG and BG stars, and the corresponding Gaussian fits are 
	shown with dotted lines. Fitted mean and standard deviations are 
	1.4\,$\pm$\,0.4\% and 2.8\,$\pm$\,1.0\%. (b) Same as panel (a) but for $\theta_{R}$ distributions of FG and BG stars. 
	Gaussian fitted mean and standard deviations are 61\,$\pm$\,11$\degr$ and 87\,$\pm$\,11$\degr$. 
	The position angle of the Galactic parallel (GP~$=$~68$\degr$) is drawn. 
	\label{fig:histogauss_p_pa}}
\end{figure*}

\begin{figure*}
        \centering
\resizebox{8.5cm}{11cm}{\includegraphics{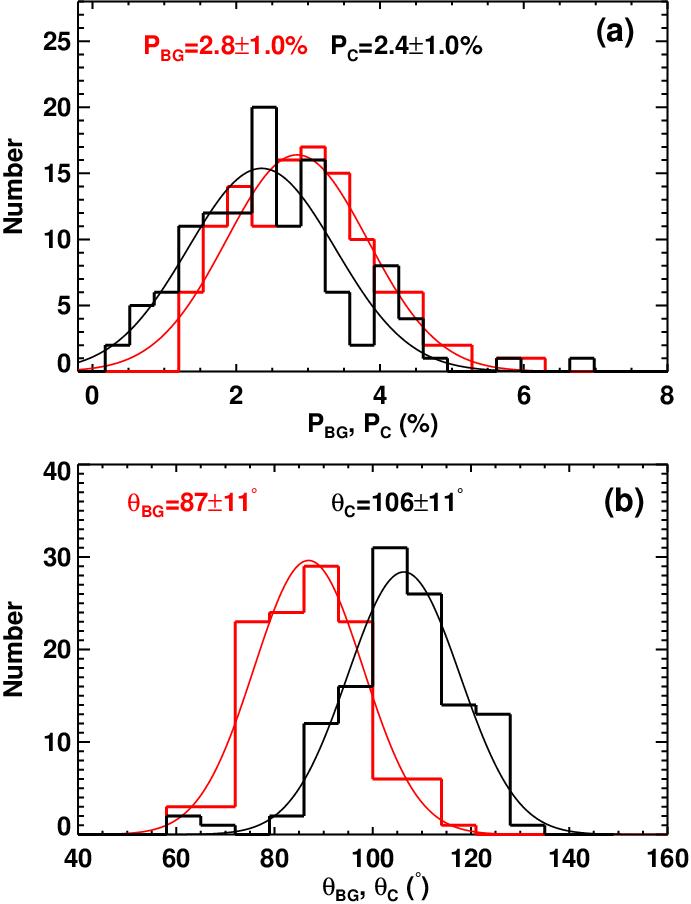}}
\resizebox{8.5cm}{11cm}{\includegraphics{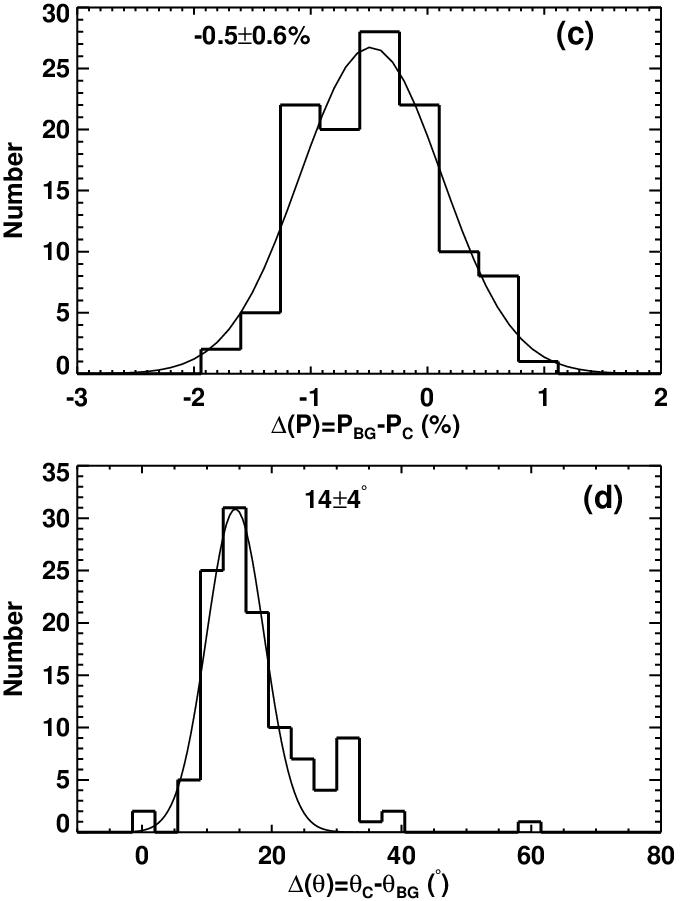}}
        \caption{(a) Number distributions of $P_{BG}$ and $P_{C}$ of the BG stars before (red histograms) and after (black histograms) ISP correction. (b) Same as (b) but for $\theta_{BG}$ and $\theta_{C}$.  (c) and (d) Distributions of $\Delta(P)~=~P_{BG}-P_{C}$ and $\Delta(\theta)~=~\theta_{C}-\theta_{BG}$, respectively. Gaussian fits and the resultant means and dispersions are shown in each panel. \label{fig:histogauss_p_pa_of_bg_c_comp}}
\end{figure*}

\begin{figure*}
        \centering
\resizebox{5.55cm}{5.55cm}{\includegraphics{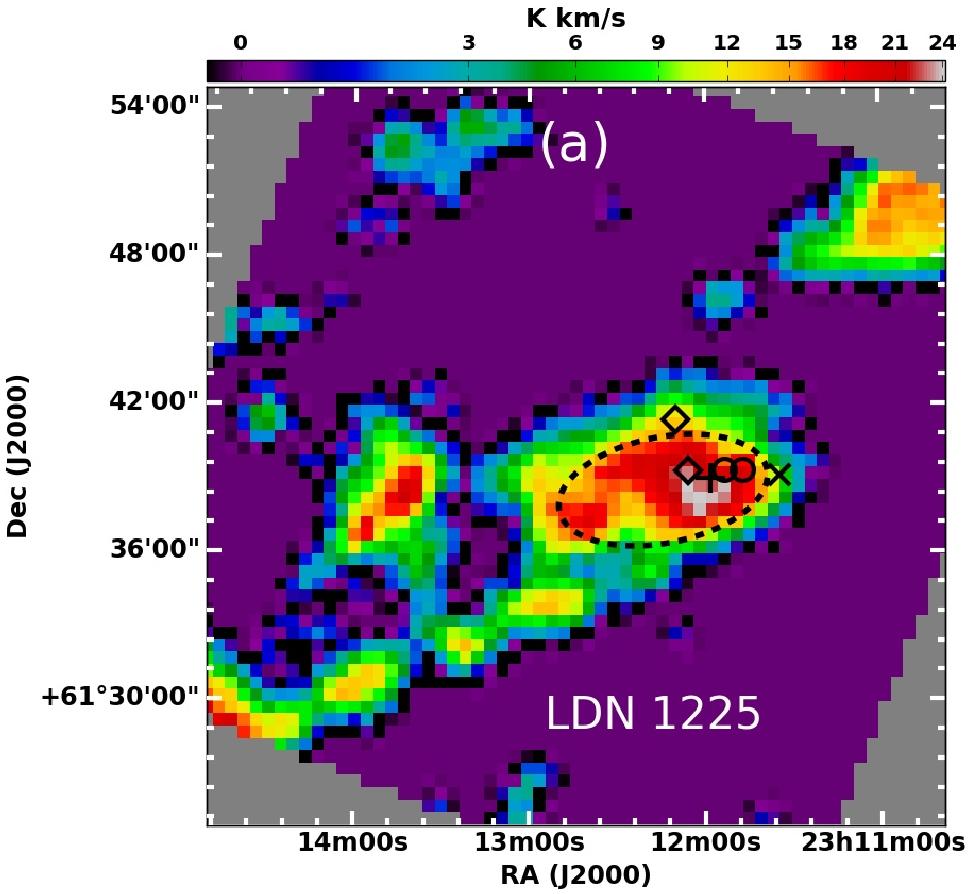}}
\resizebox{5.55cm}{5.55cm}{\includegraphics{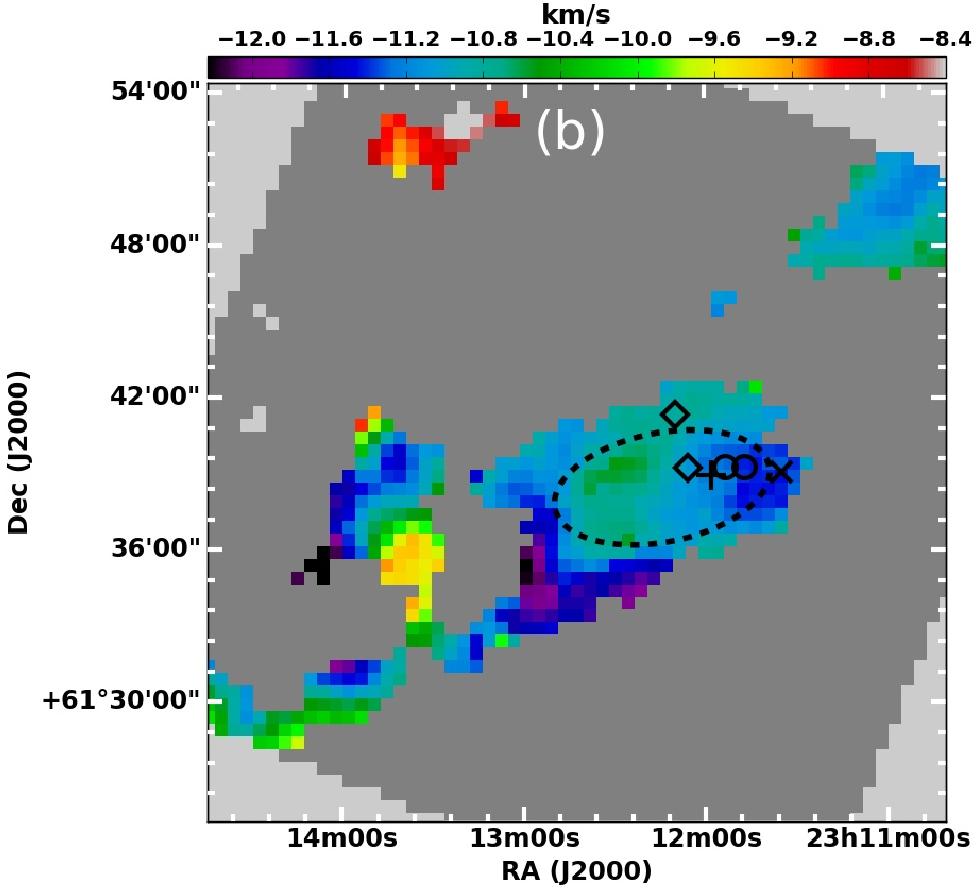}}
\resizebox{5.55cm}{5.55cm}{\includegraphics{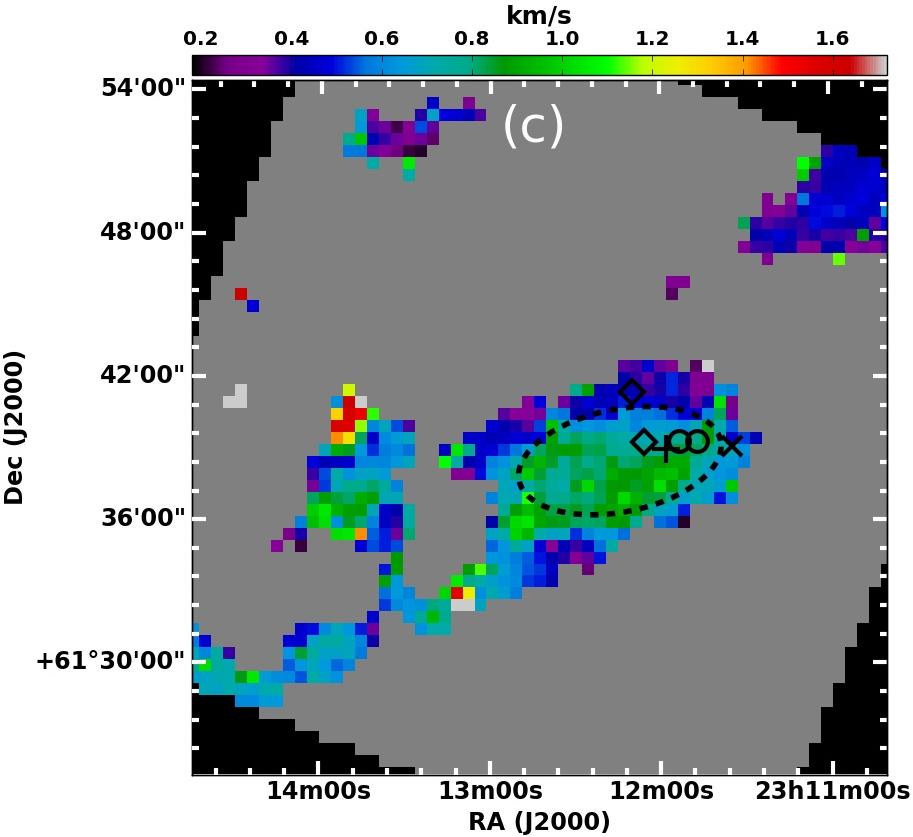}}
	\caption{$^{12}$CO (1~--~0) moment maps of the $30\arcmin\times30\arcmin$ area containing LDN\,1225 
	using the data from PMO as a part of the MWISP project. (a) CO total integrated intensity map (moment 0) 
	of the region containing LDN\,1225. (b) CO mean velocity map (moment 1). (c) CO velocity dispersion map (moment 2). 
	These maps were made using the velocity channels ranging from $\sim-$13 km s$^{-1}$ to $\sim-$3 km s$^{-1}$ 
	and having the brightness temperature above 3 times of the rms noise. Extent of the LDN\,1225 
	is shown with a dashed ellipse with major and minor axes lengths of $8\farcm6$ and $4\farcm3$, 
	and with PA of 102$\degr$. Color bar in each panel corresponds to the respective map units. Cloud center of LDN\,1225 
	is denoted with a plus symbol in all the panels. Location of IRAS 23094+6122 source is shown 
	with a cross symbol. Circles and diamonds correspond to the {\it Herschel}/PACS 70\,$\mu$m and MIR sources, 
	respectively. Within the area of LDN\,1225, the distributions of gas velocity (panel b) as well as the gas 
	dispersion (panel c) are found to be uniform. \label{fig:moments012_ldn1225}}
\end{figure*}

\begin{figure*}
\centering
\resizebox{18cm}{6cm}{\includegraphics{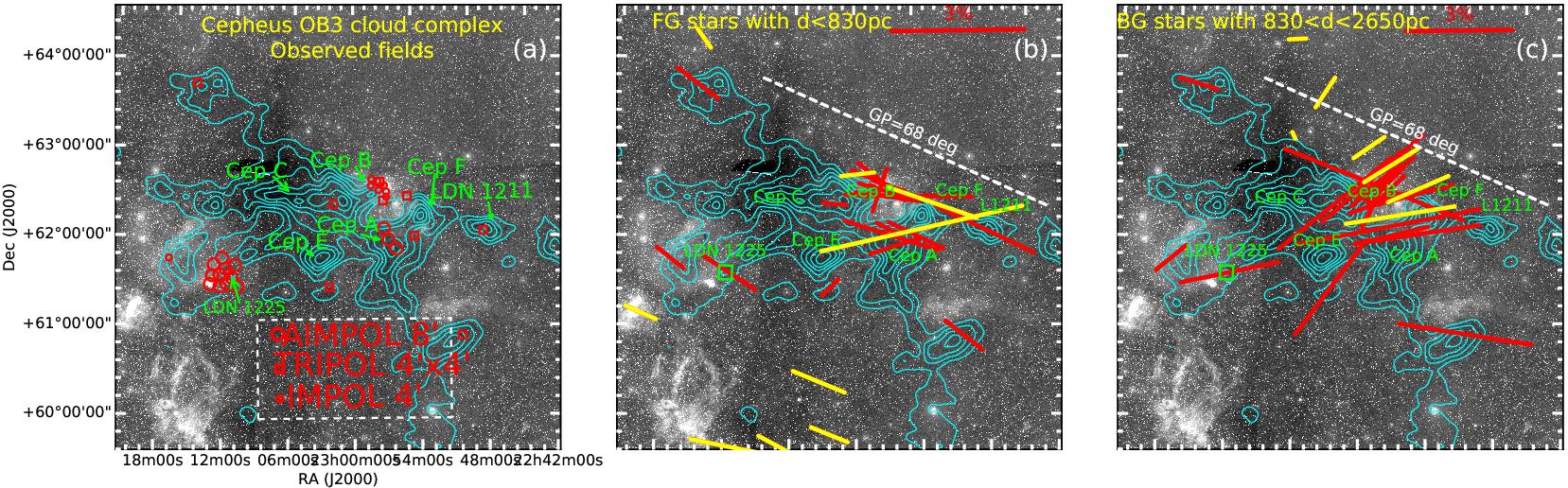}}
	\caption{(a) Polarization measurements of several fields to infer the B-field orientation towards CepOB3. Observed fields with optical polarimeters AIMPOL, TRIPOL, and IMPOL, are shown and depicted. 
	 (b) B-field orientation in the foreground medium of CepOB3 using the weighted mean polarizations of stars with $d$~$<$~830~pc. (c) Same as (b) but corresponds to the B-field map of CepOB3 using the polarizations of BG stars with 830~pc~$<$~$d$~$<$~2.65 kpc and corrected for ISP contribution. Locations of the clumps and globules are shown. Note here that in case of a field having more than two stars, their weighted means are plotted; otherwise their single measurements are shown. Similarly, we also plotted FG and BG stars (ISP corrected) polarization with \citet{Heiles2000} data using yellow vectors. In panels (a) and (b), the orientation of Galactic plane with a position angle of 68$\degr$ and a reference vector with 3$\%$ and 90$\degr$ are shown. In all the panels the background is the $5\degr\times5\degr$ DSS R-band image and the cyan contours correspond to the extinction ($A_{V}$) map drawn from 2.5 to 6.5 mag with an interval of 0.5 mag \citep{Dobashietal2005}. \label{fig:obs_fields_fg_bg_bfields_cepob3}}
\end{figure*}

\begin{figure*}
        \centering
\resizebox{13.0cm}{9.0cm}{\includegraphics{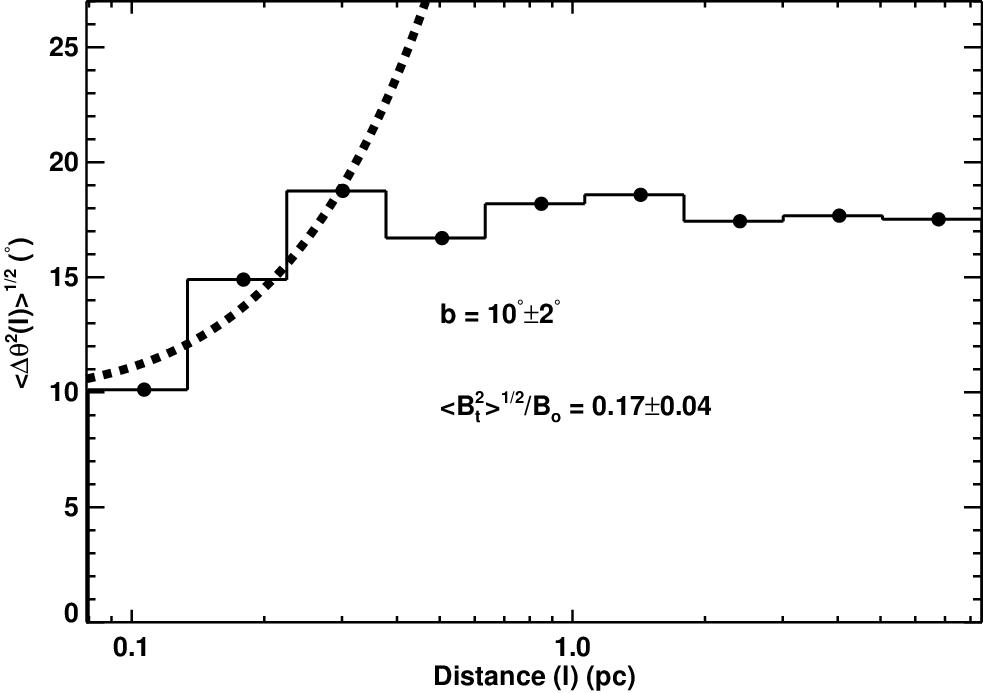}}
	\caption{Plot shows square root of the second-order structure function (or angular dispersion function (ADF)),
        ${\langle\Delta(\theta^{2}(l))\rangle}^{1/2}$ (degree) versus the distance ($l$) in pc. For this we have used
	the $\theta_{C}$ of 118 BG stars. The filled circles are the ADF values in each bin (in log scale).
        The error bars are similar to the size of the symbols. The turbulent contribution to the total angular
        dispersion function is estimated by the intercept ($b$) of the fit to the data at distance ($l$)~$=$~0.
        The measurements errors were corrected before fitting the function to the data.
        The thick dotted line corresponds to the best fit to the data for the distance less than 0.5~pc. \label{fig:structure_function}}
\end{figure*}

The CO mapping with 1.2 m telescope of the Center of Astrophysics, was carried out towards the Perseus arm 
covering the CepOB3, Cas A and NGC\,7538 \citep{Ungerechtsetal2000}. Their survey revealed that $V_{LSR}$ of the local arm 
lie between $+5$ and $-25$ km~s$^{-1}$, which is clearly separated kinematically from the distant Perseus arm whose $V_{LSR}$ lie between ~$-45$ and $-$80 km s$^{-1}$ with a mean $V_{LSR}$~$\sim-$54.5 km s$^{-1}$. It is worth stating here that there exist an emission free irregular band between $\sim-$45~km~s$^{-1}$ and $-$20~km s$^{-1}$ in the 
PV diagram \citep[][see their Figures 2a and 2b]{Ungerechtsetal2000}, suggesting a dearth of material within the inner arm. This is also true from PV cuts on $^{12}$CO data towards a small area containing LDN\,1225 as shown in Figure \ref{fig:pvcuts}. This result would also suggest the absence of an additional cloud component with a significantly different B-field orientation within the inner arm. 
Therefore, it is worth to claim that $\theta_{R}$ values lying between $\sim$70~--~$\sim$100$\degr$ (Figure \ref{fig:dist_vs_p_pa_q_u}(b)) and having distances between 830~pc~--~2.65 kpc may correspond to the B-fields of LDN\,1225, and hence the dispersion in the $\theta_{R}$ values is due to the turbulence with in the cloud.
However, a larger dispersion in the corresponding $P_{R}$ values spanning a range of $\sim$1.5~--~$\sim$5\% (Figure \ref{fig:dist_vs_p_pa_q_u}(a)) 
is owing to different optical depths, causing different amounts of polarization values as traced by the background stars lying at different parts of the cloud. 

Furthermore, the local arm possesses three gas components centered around $\sim+$5, $\sim-$5, and $\sim-$10 km s$^{-1}$. The $-$10 km s$^{-1}$ component with a relatively dense gas is attributed to the CepOB3 located at $\sim$800~pc, whereas the other two components with less dense gas may correspond to the diffuse foreground dust layers located between our Sun and CepOB3. It is interesting to state that distribution of gas with the components $-$5 km s$^{-1}$ and $-$10 km s$^{-1}$ exhibits a continuous spread implying the presence of uniformly distributed tenuous medium up to the CepOB3. 
This is further corroborated from a continuous increase in the level of polarization as a function of distance (cf., Figure \ref{fig:dist_vs_p_pa_q_u}) 
as well as from a smooth variation of $P_{R}$ with $\theta_{R}$ (cf., Figure \ref{fig:p_vs_pa}). 

Presence of two components can be witnessed in the $P$ and $\theta$ distributions of 
FG and BG stars (Figure \ref{fig:histogauss_p_pa}, sections \ref{subsec:fgbgidentification} 
and \ref{subsec:fg_subtraction}). 
B-field component of the FG medium is oriented along the Northeast--Southwest 
(with a mean $\theta_{FG}$~$=$~61\,$\pm$\,11$\degr$), while that of the cloud medium is along 
East--West (with a mean $\theta_{C}$~$=$~106\,$\pm$\,11$\degr$).
These two components, with different mean polarization values, 
are separated at $P_{R}$~$=$~2\% and $\theta_{R}$~$=$~75$\degr$ (Section \ref{subsec:distriofppa}). 
In summary, our polarization observations revealing two distributions in both degrees of polarization and polarization angles are 
consistent with two CO cloud components centered around $\sim-$5~km~s$^{-1}$ and $\sim-$10~km~s$^{-1}$. 

\subsection{Geometry of LDN\,1225~in $^{12}$CO, $^{13}$CO, and C$^{18}$O molecular lines}\label{districolines}

Generally the $^{12}$CO emission comes from the tenuous gas distributed in low density parts of the cloud, whereas the $^{13}$CO and C$^{18}$O emissions come from the 
inner, denser gas. Since LDN\,1225 has been observed simultaneously at these three molecular lines, we compare the ambient mean B-fields orientation with the orientations of major axes of LDN\,1225 at different length scales and depths. These comparisons 
would reveal how the B-fields, turbulence, and gravity interact with the cloud material and 
govern its structure, stability, and contraction at different scales and depths \citep{Eswaraiahetal2013}. 
The advantage in using molecular lines to determine the cloud structure is that 
the foreground cloud component will be well separated from that of the background in velocity space. 

\begin{figure*}
        \centering
\resizebox{8.8cm}{7.75cm}{\includegraphics{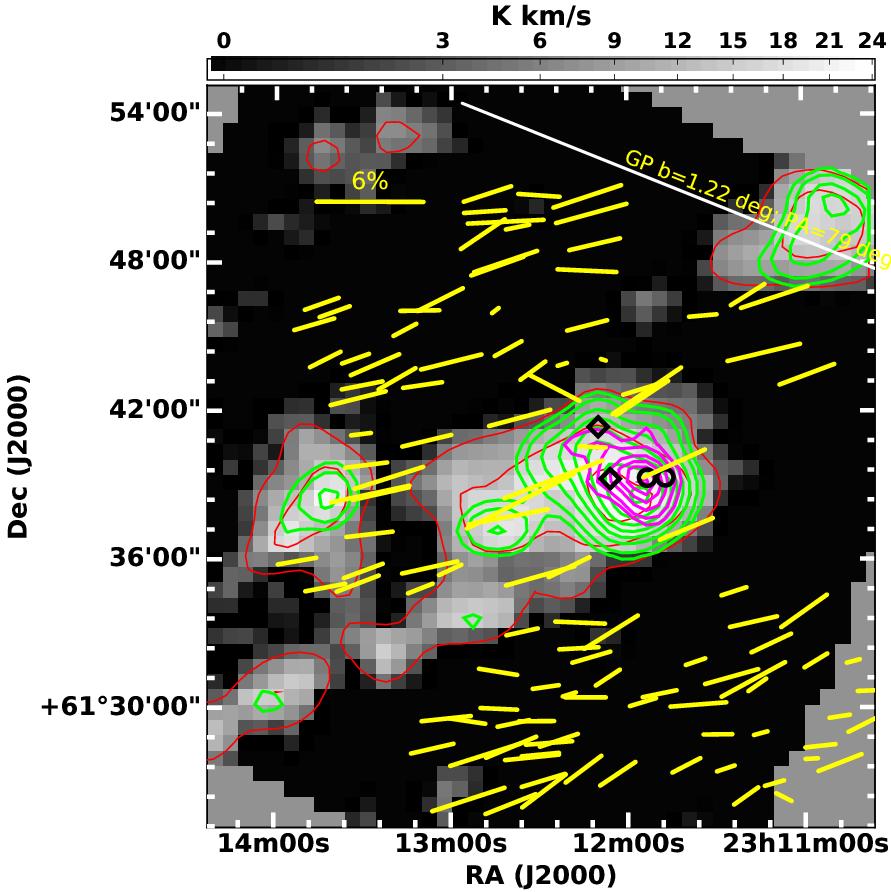}}
	\caption{Structure of LDN\,1225 as traced by $^{12}$CO, $^{13}$CO, and C$^{18}$O molecular lines.
Background image is the moment 0 map of $^{12}$CO and the corresponding contours are shown with red at 3, 12, and 21 K km s$^{-1}$. $^{13}$CO moment 0 contours are shown with green at 0.5, 1, 2, 3, 4, 5, 6, and 7 K km s$^{-1}$. C$^{18}$O moment 0 contours are shown with magenta at 0.075, 0.15, 0.30, 0.45, 0.60, 0.75, and 0.9 K km s$^{-1}$. Polarization vectors are same as in Figure \ref{fig:polvecmap_colorcompo}(b). Two 70\,$\mu$m and two MIR sources are shown with circles and diamonds, respectively. \label{fig:structure_LDN1225_colines}}
\end{figure*}

We have employed CASA 2D Gaussian fitting function on the moment 0 maps of $^{12}$CO, $^{13}$CO, and C$^{18}$O, respectively shown with red, green, and magenta contours in Figure \ref{fig:structure_LDN1225_colines}, and 
extracted the spatial extents and position angles of the cloud elongations. 
The corresponding fitted central coordinates, 
FWHMs of major and minor axes, and position angles are given in Table \ref{tab:cloudpas}. 
Last column gives the offset ($\Delta(\theta)$) between the cloud position angles ($\theta_{cloud}$) traced by 
three CO lines and mean B-field orientation ($\theta_{C}$, Section \ref{subsec:fg_subtraction}) of LDN\,1225. 

The $\Delta(\theta)$ from $^{12}$CO is 4$\pm$11$\degr$, implying 
low density outer parts of the cloud are nearly aligned parallel with the ambient B-fields, 
suggesting importance of B-fields in governing the cloud structure. 
However, $\Delta(\theta)$ values from $^{13}$CO and C$^{18}$O to be 54\,$\pm$\,11$\degr$ and 52\,$\pm$\,14$\degr$; indicating that, within the error, the PAs of the denser parts of the cloud are neither parallel nor perpendicular to B-fields. This suggests that B-fields of the low-density outer parts may not equally be important in governing the cloud structure at high-density inner parts; in other words turbulence and gravity might be crucial at these denser parts. In order to shed more light on these, we need to probe the B-fields  in the denser parts, using NIR and sub-mm polarimetry. 

\subsection{Association of MIR and 70\,$\mu$m point sources with LDN\,1225}\label{subsec:70micronsources}

\begin{figure*}
        \centering
        \resizebox{8.8cm}{7.75cm}
{\includegraphics{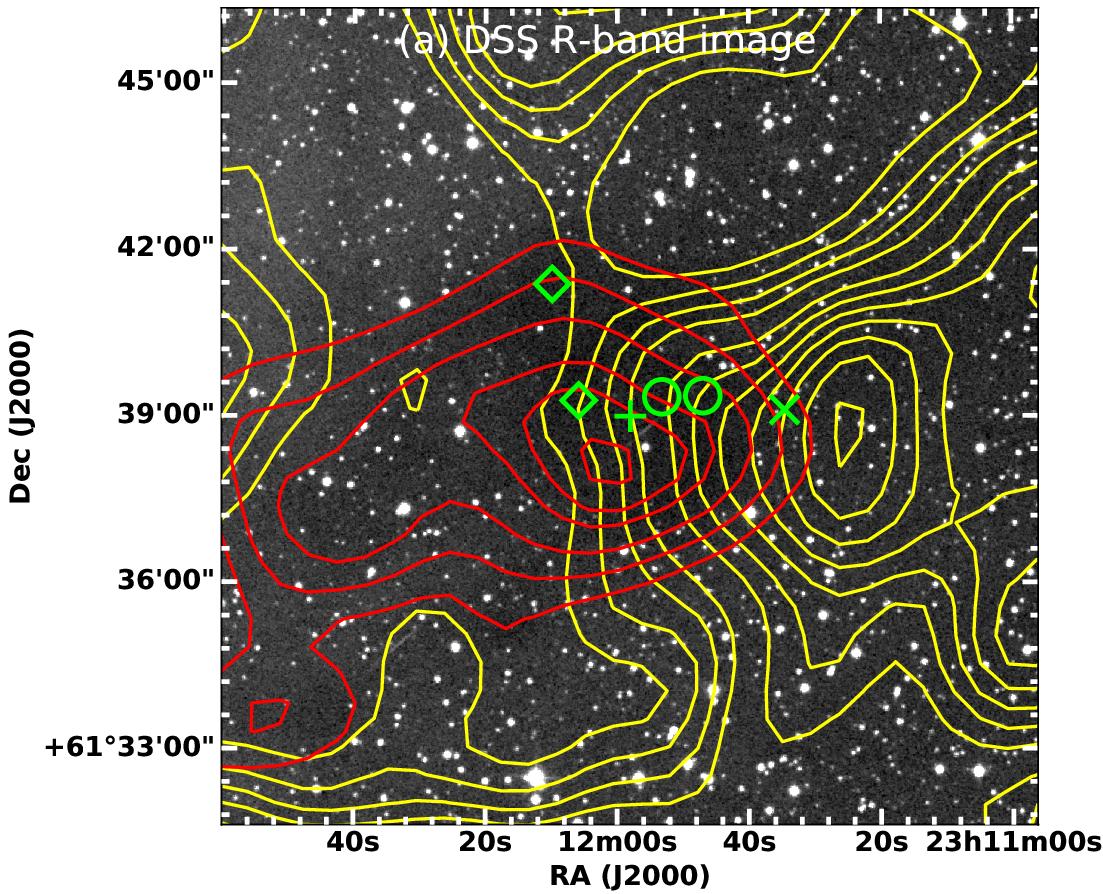}}
\resizebox{8.8cm}{7.5cm}
{\includegraphics{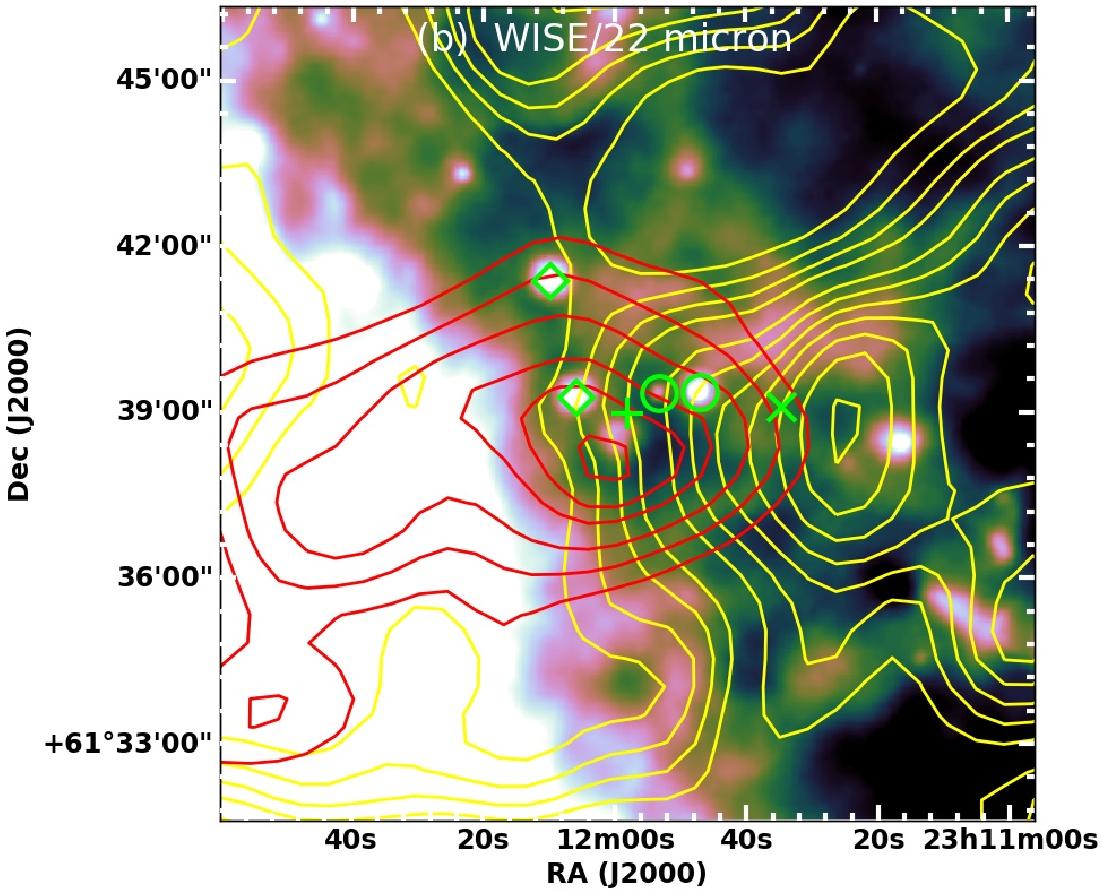}}
\resizebox{8.8cm}{7.5cm}{\includegraphics{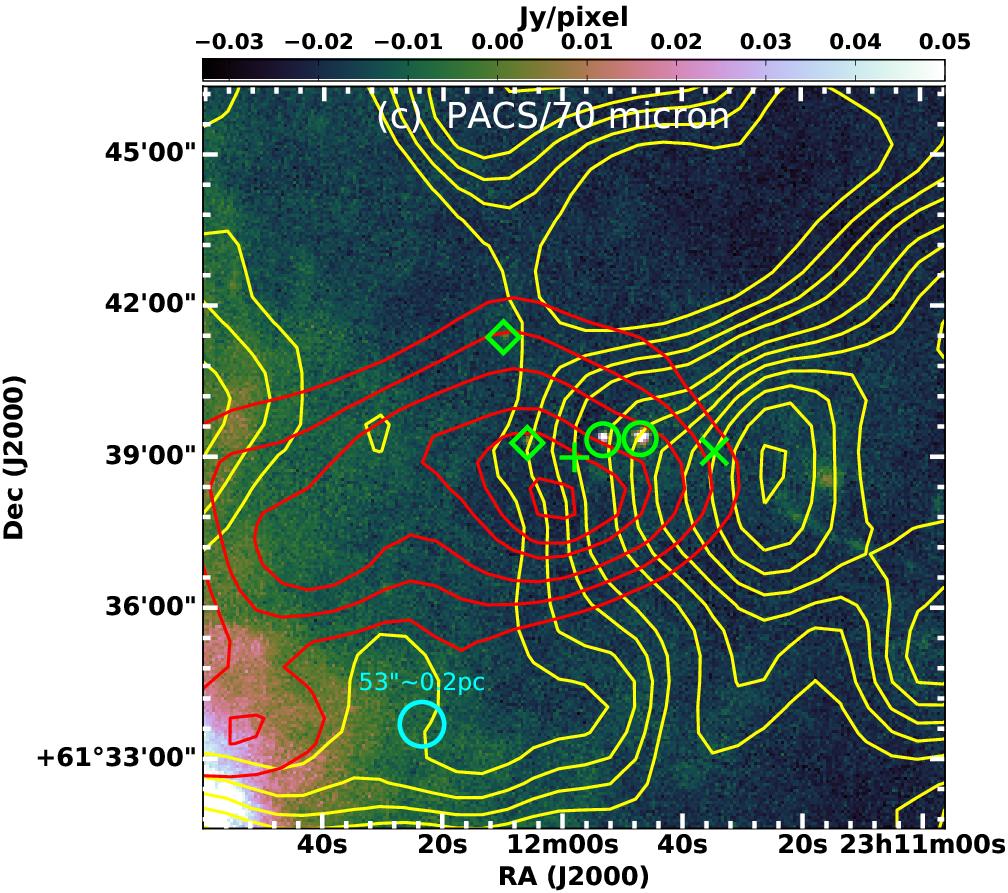}}
\resizebox{8.8cm}{7.75cm}{\includegraphics{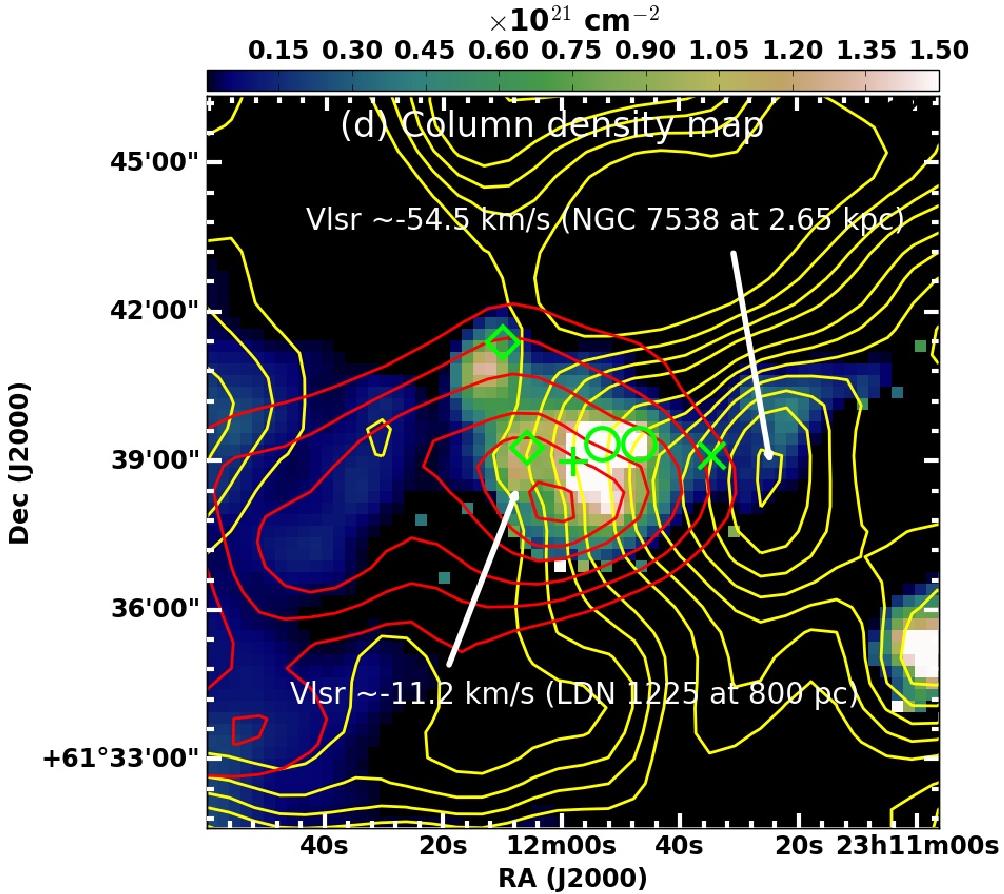}}
\caption{Locations of the 70\,$\mu$m sources (circles) and the distributions of moment 
	0 maps of two $^{12}$CO gas components overlaid on multi-wavelength images: 
	(a) DSS R-band image, (b) WISE 22\,$\mu$m image, (c) {\it Herschel}/PACS 70\,$\mu$m
image, and (d) 
column density map. Red contours, drawn at 6, 10, 14, 18, 20, and 22 K km s$^{-1}$, 
	correspond to $^{12}$CO moment map of LDN\,1225 with mean $V_{LSR}$~$-$11.2 km s$^{-1}$. 
	Yellow contours, drawn at 6, 10, 14, 18, 25, 30, 35, 40, 45, 50, and 55 K km s$^{-1}$, 
	correspond to $^{12}$CO moment map of distant star forming region in Perseus arm with mean $V_{LSR}$~$-$54.5 km s$^{-1}$. 
Two bright MIR sources, denoted with diamond symbols, are faint in 70\,$\mu$m image. Cloud center of LDN\,1225 is 
	denoted with a plus symbol. Location of IRAS 23094+6122 source is shown with a cross symbol. 
	Beam size (53$\arcsec\simeq$~0.2~pc) of the CO map is shown with a cyan circle in panel (c). 
	The centers of the CO moment 0 emission peaks of two cloud components are shown with arrows in panel (d). 
	Color bar in panels (c) and (d) correspond to the flux units of the respective image. At 22\,$\mu$m image 
	(panel b), the SE region of LDN\,1225 has a significant amount of contamination from the background star-forming region NGC\,7538. \label{fig:twococomponents}}
\end{figure*}

In order to confirm if LDN\,1225 is a starless or star forming cloud, 
we search for the YSOs by examining the multi-wavelengths images of LDN\,1225 
as shown in the Figure \ref{fig:twococomponents}. 
We have identified two MIR (diamonds) and two 70 $\mu$m (circles) sources, respectively, 
in the images of {\it WISE} 22 $\mu$m 
(Figure \ref{fig:twococomponents}(b)) and {\it Herschel}/PACS 70\,$\mu$m 
(Figure \ref{fig:twococomponents}(c)). These sources appear to be part of LDN\,1225, because 
as they are located within the extent and center of the cloud as seen in optical 
absorption (Figure \ref{fig:twococomponents}(a)) 
and dust emission inferred from the column density map\footnote{Column density map has 
been constructed by fitting the modified blackbody function on the {\it Herschel} PACS and SPIRE fluxes as 
described in Appendix D of \citet[][and references therein]{Eswaraiahetal2017}. Data have been 
take from {\it Herschel} Infrared Galactic Plane Survey \citep[Hi-GAL;][]{Molinarietal2010}.} (Figure \ref{fig:twococomponents}(d)). 
Nonetheless, CO data cubes reveal two overlapping cloud components 
towards LDN\,1225 -- one at $V_{LSR}$~$=$~$-$11.2~km~s$^{-1}$ (red contours) 
and the another at $V_{LSR}$~$=$~$-$54.5~km~s$^{-1}$ (yellow contours) as shown in Figures \ref{fig:twococomponents}(a)~--~(d) (see also PV cuts in 
Figures \ref{fig:pvcuts}(a) and (b)). 
Because of this, a confusion prevails on whether YSOs are originally belong to the LDN\,1225 or to the background cloud NGC\,7538. 

\begin{figure*}
        \centering
\resizebox{5cm}{5.5cm}{\includegraphics{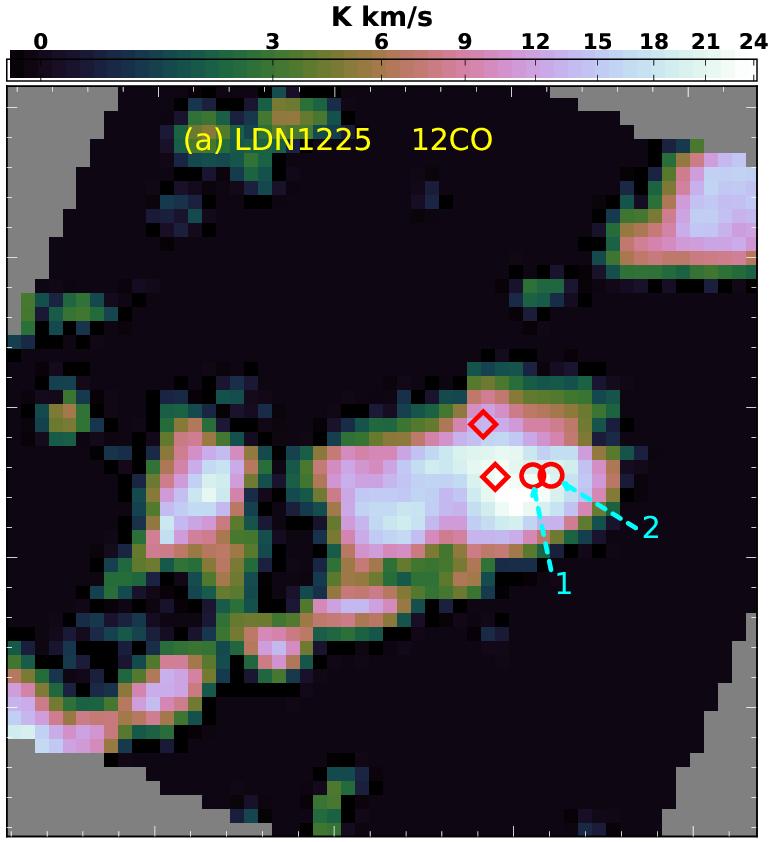}}
\resizebox{5cm}{5.5cm}{\includegraphics{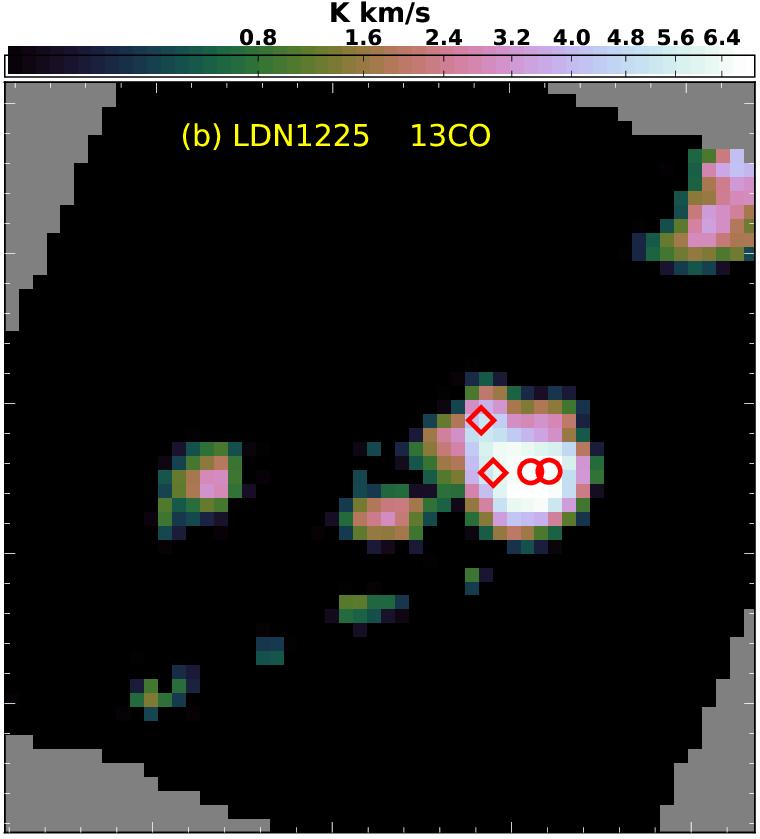}}
\resizebox{5cm}{5.5cm}{\includegraphics{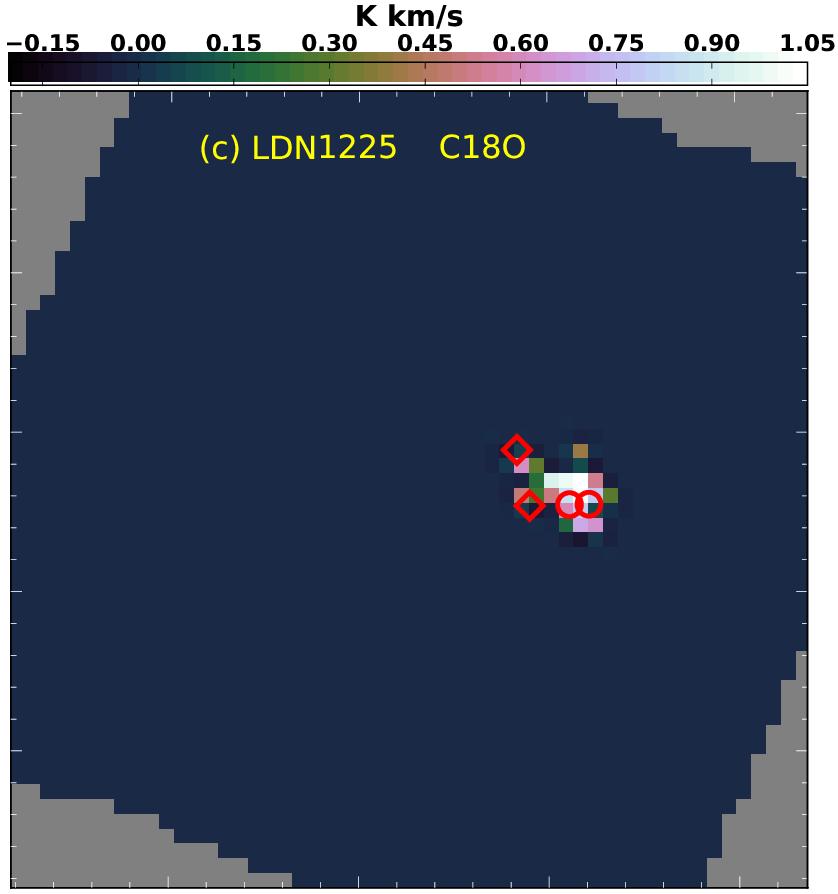}}
	\resizebox{5cm}{5.5cm}{\includegraphics{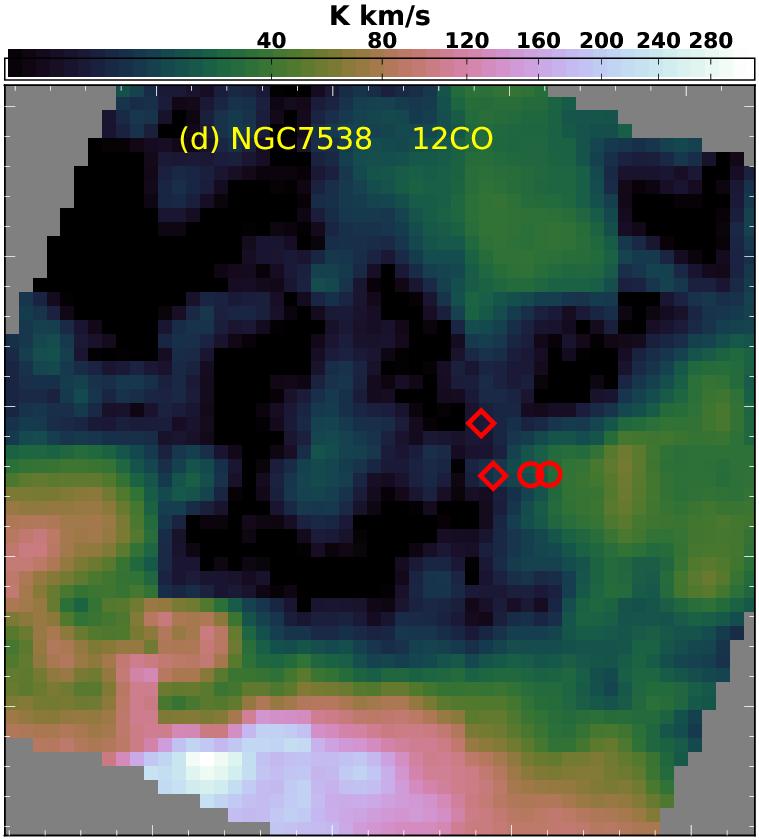}}
	\resizebox{5cm}{5.5cm}{\includegraphics{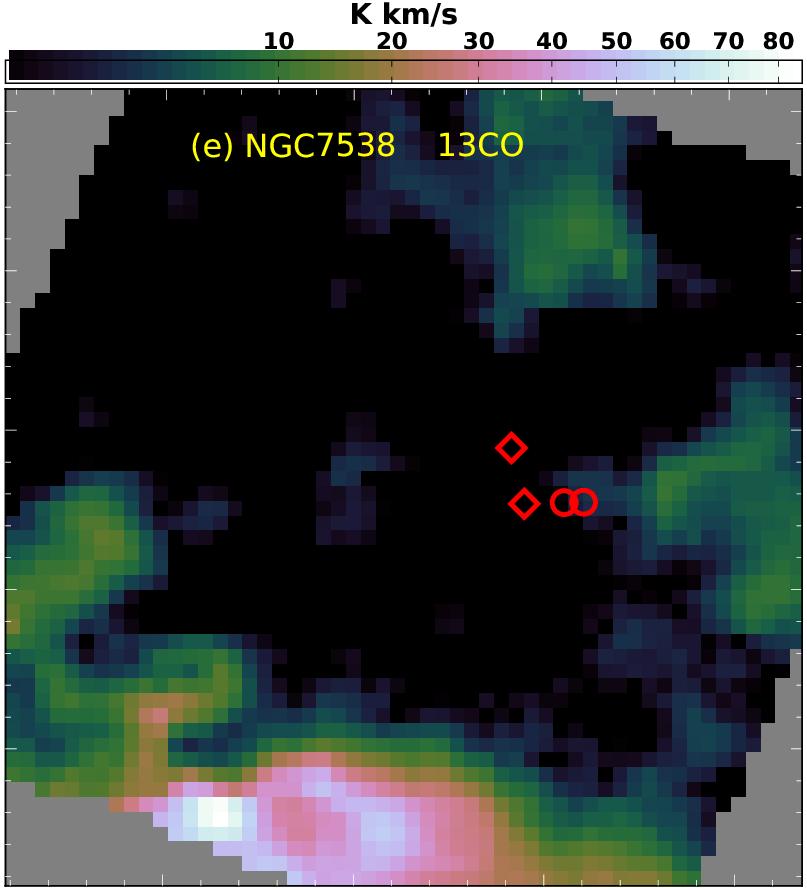}}
        \resizebox{5cm}{5.5cm}{\includegraphics{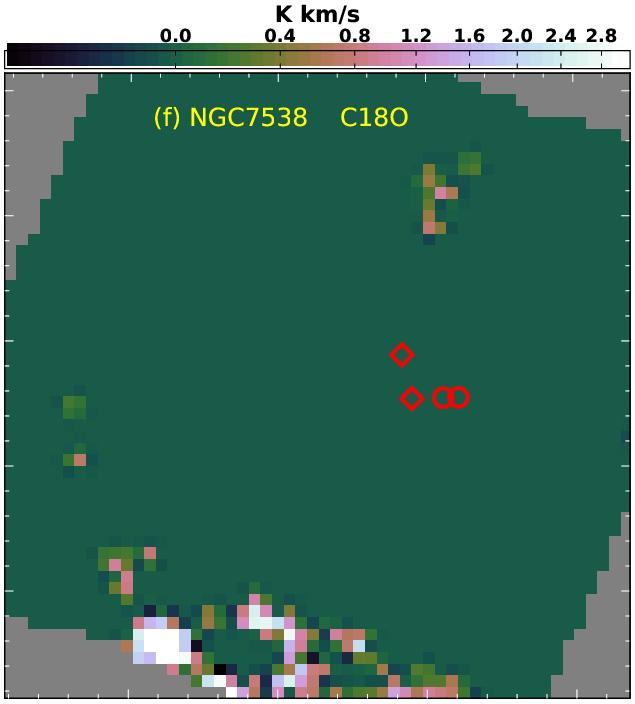}}
	\caption{ The $^{12}$CO (left),  $^{13}$CO (middle), and  C$^{18}$O (right) moment 0 maps of the LDN\,1225 (panels (a), (b), and (c)) 
	and background/NGC\,7538 cloud (panels (d), (e), and (f)). Locations of MIR (diamonds) and 70\,$\mu$m (circles) sources are shown. 
	Cyan arrows in panel (a) represent the id numbers of 70\,$\mu$m sources. Color bar in each panel corresponds to the pixel values (K km s$^{-1}$). Coordinates and dimensions of the maps are same as that of Figure \ref{fig:structure_LDN1225_colines}. North is up and east is to the left. \label{fig:moment0_maps_ldn1225_background}}
\end{figure*}

To elucidate more on the association of YSOs, we have searched for the $^{12}$CO, $^{13}$CO, and C$^{18}$O emissions from
LDN\,1225 (top panels) and NGC\,7538 (bottom panels), especially, at the locations of the YSOs as shown in Figure \ref{fig:moment0_maps_ldn1225_background}.
Evidently, all YSOs are distributed, not only within the emissions from 
low-density tracers ($^{12}$CO/$^{13}$CO) but also that from dense gas tracer (C$^{18}$O) of LDN\,1225. 
Moreover, despite of a small amount of $^{12}$CO/$^{13}$CO emission, 
a complete devoid of C$^{18}$O emission from NGC\,7538 is witnessed at the sites of YSOs. This implies that LDN\,1225 comprises of denser gas than 
background cloud. These results reinforce the fact that LDN\,1225 is indeed a star forming dark globule, hosting MIR and 70\,$\mu$m sources in its denser parts.

\subsection{Properties of 70\,$\mu$m sources}\label{70micronprop}

\begin{figure*}
        \centering
        \resizebox{8.5cm}{6.25cm}{\includegraphics{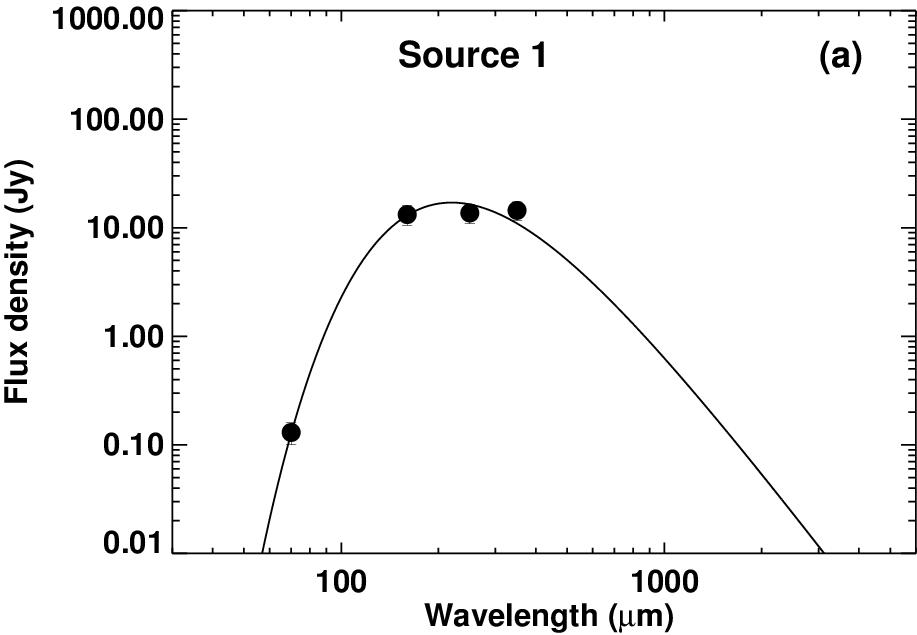}}
        \resizebox{8.5cm}{6.26cm}{\includegraphics{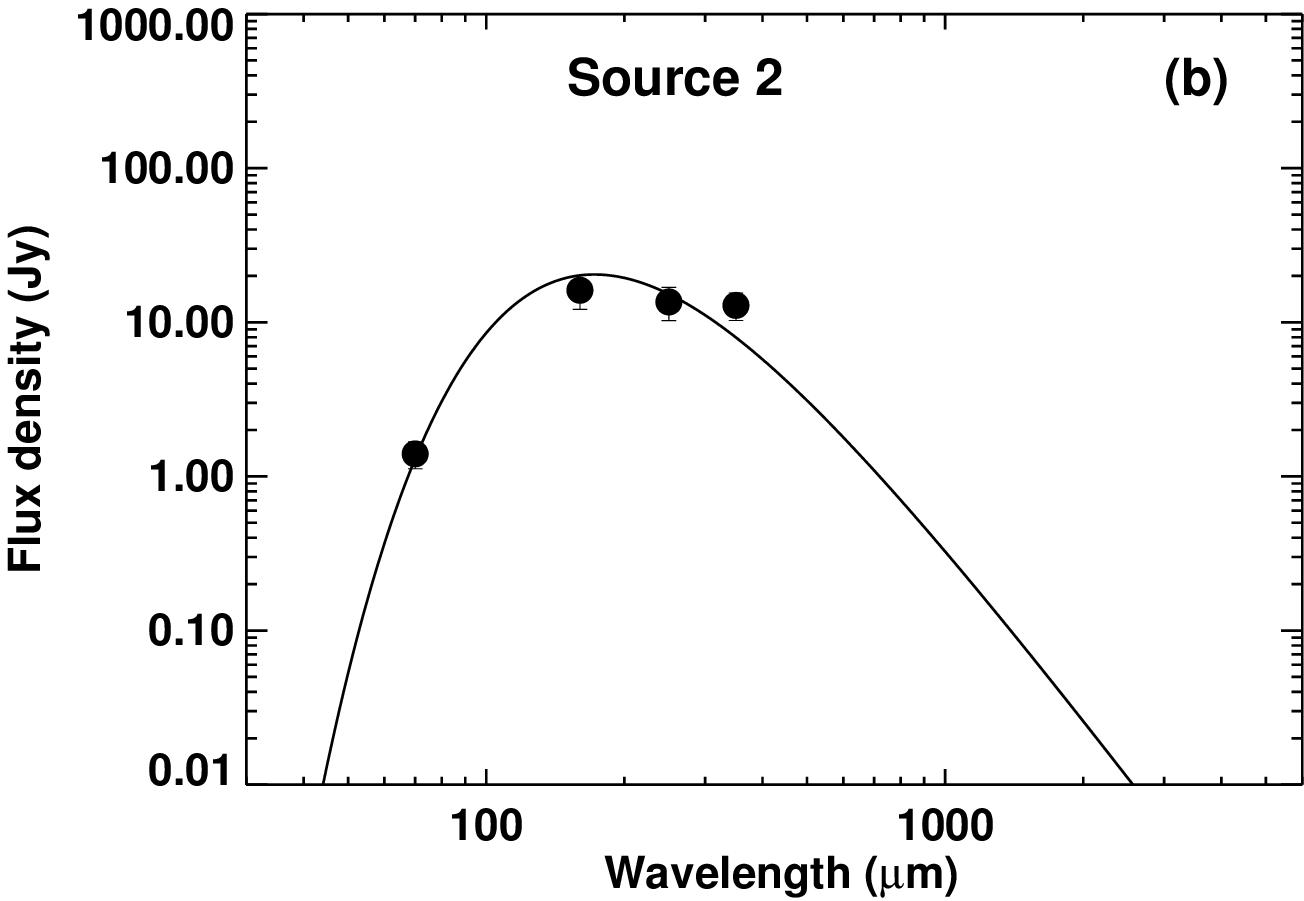}}
	\caption{The SEDs of envelopes of two 70\,$\mu$m sources. 
	Black line shows the best modified blackbody fit to the input fluxes (filled circles) between 70~--~350
	$\mu$m.  Panels (a) \& (b) correspond to the sources 1 \& 2, respectively (id numbers 
	are shown in Figure \ref{fig:moment0_maps_ldn1225_background}(a)). \label{fig:seds70micronsources}}
\end{figure*}

Since 70\,$\mu$m sources are brighter in FIR and fainter in NIR and MIR wavelengths  
(cf., Figure \ref{fig:twococomponents}), their disks may have negligible contribution to 
the shape of spectral energy distribution (SED) at $\lambda$~$\leq$~100 
$\mu$m \citep[eg.,][]{Whitneyetal2005}. The main contributor to the SED of 70\,$\mu$m sources 
could be from their envelopes. In order to study the nature of these 
70\,$\mu$m sources (cf., Section \ref{subsec:70micronsources} and Figure \ref{fig:moment0_maps_ldn1225_background}), 
we have performed aperture photometry on the {\it Herschel} images as documented in \citet{Balogetal2014} and 
modified blackbody fitting was performed on the fluxes at 70, 160, 250, and 
350\,$\mu$m\footnote{Fluxes at 500\,$\mu$m have been excluded, owing to the low-resolution of the 500\,$\mu$m beam. 
This is done in order to avoid the baisness in the fitting procedure due to overestimation of fluxes 
at 500\,$\mu$m because of source confusion and inclusion of excess background emission.}  as 
shown in Figure \ref{fig:seds70micronsources}(a) and (b). 

The computed envelope masses ($M_{env}$), bolometric luminosities ($L_{bol}$), 
and dust temperatures ($T_{dust}$) are 11 M$_{\sun}$ \& 4 M$_{\sun}$, 
5 L$_{\sun}$ \& 8 L$_{\sun}$, and 13 K \& 17 K, respectively, 
for sources 1 \& 2. SED fits suggest that the envelopes of the two 70\,$\mu$m sources 
are of low-mass and low-luminous Class 0 protostars.

\begin{figure*}
        \centering
        \resizebox{8.5cm}{7.0cm}{\includegraphics{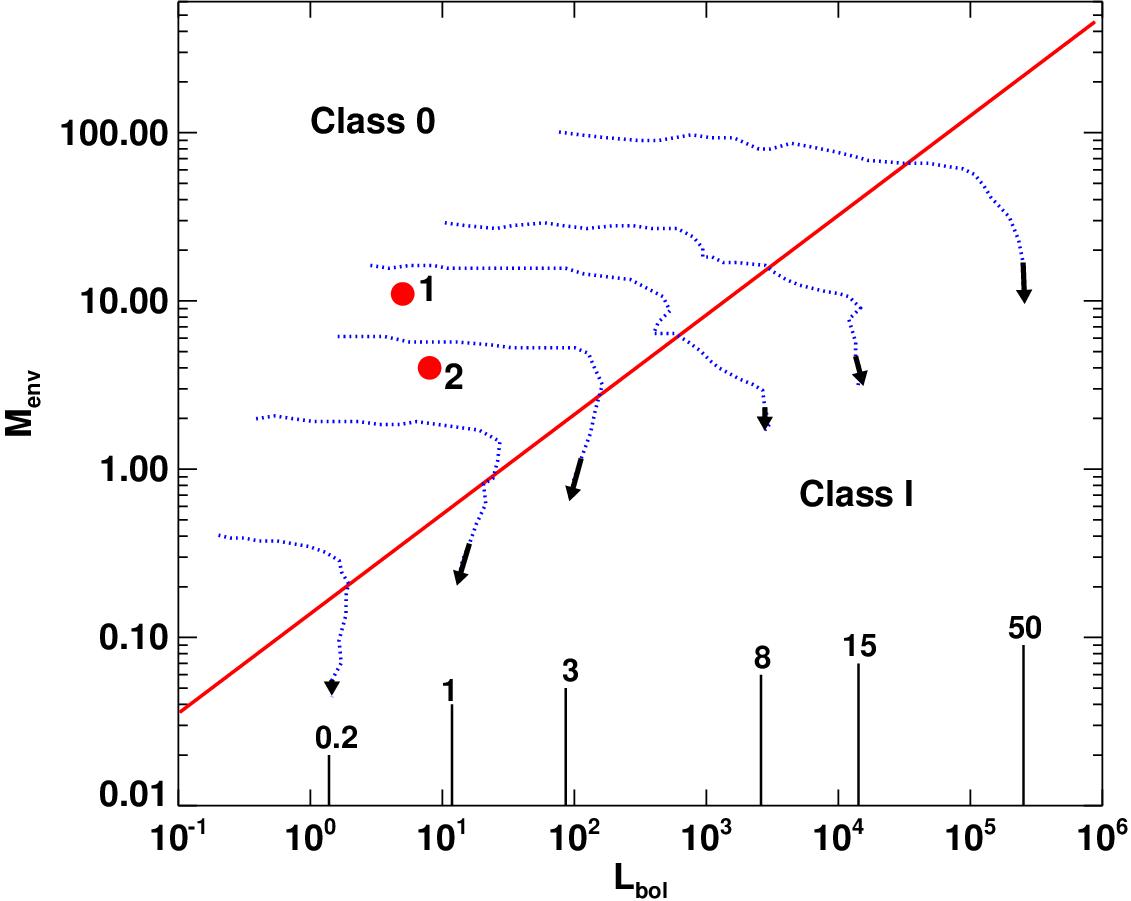}}
        \caption{Bolometric luminosity (L$_{bol}$) vs. envelope mass (M$_{env}$) for the two 70$\mu$m sources (filled circles with IDs). The dotted blue lines represent the evolutionary tracks from \citet{Andreetal2008}. Evolution proceeds from the upper left to the lower right (depicted with arrows at the end of each track). The final stellar masses of these tracks in solar units are given above the lower axis. The slanted red line corresponds to the location where 50\% of the initial core mass is converted into stellar mass \citep[see][]{Bontempsetal1996,Andreetal2000}.  \label{fig:LumMass_70micronsources}}
\end{figure*}

Furthermore, we have used the above physical parameters of these sources to infer 
their final evolutionary status based on the $M_{env}$--$L_{bol}$ diagram (Figure \ref{fig:LumMass_70micronsources}). 
The location of the two 70\,$\mu$m sources in comparison to the evolutionary tracks of protostars 
with different envelope masses and luminosities \citep[][and references therein]{Andreetal2008}, suggest that 
they may evolve into stars with their final masses span over 1 M$_{\sun}$ to 5 M$_{\sun}$ at the end phases of accretion 
(see the tip of the arrows in Figure \ref{fig:LumMass_70micronsources}).

 
 \subsection{Are the B-fields important to the Star formation in LDN\,1225?}
 
 If B-fields are important in a cloud to form the cores 
 via ambipolar diffusion process or strong B-field model 
 \citep{Mouschovias1976,Shu1977,Shuetal1987,Mouschovias1991,MouschoviasCiolek1999} 
 then (i) the B-field structure in the low density subcritical envelope 
 should be aligned with the cloud structure, (ii) high-density core's major axis 
 should be perpendicular to the envelope B-fields, and (iii) 
 B-fields within the core should follow an hour-glass shape if the core is in a supercritical state and collapsing. 

 The first condition is true in LDN\,1225 because 
 the mass-to-magnetic flux ratio in units of the critical value, $\mu$, is $<$~1 suggesting 
 that envelope is in a subcritical state 
 (section \ref{subsec:cloud_stability}) and hence the low-density envelope is strongly supported by B-fields. 
 In addition, being parallel to the low-density parts (section \ref{subsec:fg_subtraction}), 
 B-fields aid the cloud contraction producing elongated core structures perpendicular to them. 
However, since the core long axis is neither parallel nor perpendicular to the 
ambient B-fields (section \ref{districolines}), the second condition may not be true in LDN\,1225.

 
High density cores are in a super-critical state as they are collapsing to form stars (Section \ref{subsec:70micronsources}). 
Since we have not probed the B-fields at the core scale and 
have not derived various parameters (B-field strength, B-field pressure, turbulent pressure, mass-to-flux ratio in terms of 
 critical value, etc), the examination of the strong B-field model (third condition) is beyond the scope of this work. 
 Therefore, to shed more light on whether or not the B-fields are important at core scales, we need to 
 probe the B-fields using NIR/sub-mm polarimetry.

\subsection{Extinction law}\label{subsec:extlaw}

Dust grain mean size distribution can be evaluated by using the parameter $R_{V}$~$=$~$A_{V}/E(B-V)$, the total-to-selective extinction which is also termed as extinction law \citep{Cardellietal1989,OlofssonOlofsson2010}. Though, the mean $R_{V}$ for the Milky Way Galaxy is 3.1, it varies from one line of sight to the other.
To characterize the extinction law towards LDN\,1225, we have used two-color diagrams (TCDs) of the form $(V-\lambda)$ versus $(B-V)$, where $\lambda$ is the one of the magnitude in broad-band filters, $R$, $I$, $J$, $H$,  $K_{s}$, or $L$. These diagrams are usually employed to characterize the extinction law in the foreground and cloud (or stellar cluster) mediums
\citep[eg.,][]{Pandeyetal2000,Pandeyetal2003,Samaletal2007,Eswaraiahetal2013,Pandeyetal2013,Kumaretal2014a}. Since the distance of LDN\,1225 and the number of FG/BG stars are known, as illustrated in Appendix \ref{subsec:rv_estimation}, we derive the $R_{V}$ values for the foreground and cloud mediums. For this purpose we have used the optical and NIR photometric data from Table \ref{tab:phot689stars}. 

Based on the two-color diagrams (Figures \ref{fig:TCD_rv_estimation1} and \ref{fig:TCD_rv_estimation2}) and 
estimated weighted mean $R_{V}$ values (see Appendix \ref{subsec:rv_estimation} for more details on the Figures and Table), the following points can be made:
(a) FG and BG stars of groups I (stars with photometry) and II (stars with photometry plus polarimetry) exhibit conspicuously two different distributions represented by different amount of slopes and $R_{V}$ values, (b) within the error, $R_{V}$ values of FG (BG) stars belong to the two groups are nearly same and consistent with each other, (c) the extinction law in the foreground medium is normal i.e., $R_{V}\sim$3.1 characterized with mean dust grain sizes, and (d) the extinction law in the cloud medium is abnormal i.e., $R_{V}\sim$3.4 characterized with relatively bigger dust grain sizes in the regions of LDN\,1225 that we probed with photometry and polarimetry. 

Different dust size is found towards different Galactic line of sights. For example, toward the high-latitude translucent molecular cloud direction of HD\,210121 \citep{WeltyFowler1992,Larsonetal1996}, $R_{V}$ is 2.1, whereas toward the HD\,36982, molecular cloud direction in the Orion nebula, the $R_{V}$ values lie between 5.6~--~5.8  \citep{Cardellietal1989,Fitzpatrick1999,Draine2003}. In our work, an abnormal extinction law (with $R_{V}\sim$ 3.4) in LDN\,1225 might have produced by the bigger dust grains resulted via dust grain growth by means of dust coagulation or dust accretion processes. These processes could take place in the molecular clouds because of the prevailing different physical conditions such as low temperatures and high density. 

  
\subsection{Extinction versus polarization and polarization efficiency}\label{subsec:ext_vs_pol_poleffi}

To understand the polarization efficiency of the dust grains of LDN\,1225, as illustrated in Appendix \ref{subsec:estim_av}, we estimate the visual extinctions ($A_{V}$) of the field stars (mainly normal main-sequence stars and giants) of LDN\,1225 by dereddenining their NIR colors (Table \ref{tab:phot689stars}) shown in Figure \ref{fig:nir_ccd}.
Figure \ref{fig:polefficiency}(a) shows the $A_{V}$ versus $P_{R}$ plot of 27 stars (of these 3 FG and 24 BG stars). 
Polarization values of all the stars (except 2 BG stars) distributed below the observed upper limit polarization relation $P/A_{V}$~$<$~3\% mag$^{-1}$ \citep{Serkowskietal1975} and exhibit a linear trend as a function of extinction. 

The polarization efficiency ($P_{R}/A_{V}$) of the dust grains of LDN\,1225 tends to follow a systematically declining trend as a function of extinction ($A_{V}$), similar to the other works \citep[e.g.,][]{Goodmanetal1995,Gerakinesetal1995,Whittetetal2008,Eswaraiahetal2013,Wangetal2017}, by following a power-law of the form $P/A_{V}\propto{A_{V}}^{-\alpha}$ shown in Figure \ref{fig:polefficiency}(b). 
Weighted power-law fit to the data points of 22 BG stars (3 FG and 2 BG stars are excluded from the fit; see figure caption for more details) satisfies the relation $P_{R}/A_{V}$~$\propto$~${A_{V}}^{-0.7\,\pm\,0.5}$. 
Within the error, the power-law index $-$0.7\,$\pm$\,0.5 is 
consistent with the index $-$0.7\,$\pm$\,0.1 for the entire molecular cloud complex IC\,5146 using $R_{c}$ band data \citep{Wangetal2017}, $-$0.8\,$\pm$\,0.1 for Pipe-109 using $R$-band data \citep{Alvesetal2014}, and $-$0.7\,$\pm$\,0.1 for L2014 using $H$-band data \citep{CashmanClemens2014}. These reported indices are generally found for the low-density parts of the cloud observed in optical wavelengths, having $A_{V}$ values of 1~--~4 mag. However, softer indices ($-$0.3 to $-$0.5) are found towards the dense parts of the clouds with high extinction regions ($A_{V}$~$>$~5 mag) observed in NIR wavelengths 
\citep{Whittetetal2008,Jonesetal2015,Wangetal2017}. 

The negative index $-$0.7\,$\pm$\,0.5 implies that dust grains at the low-density outer parts result higher polarization efficiency, while those at relatively high-density inner parts result lower polarization efficiency. This is due to the variation of several factors as a function of extinction \citep{Jones1989,Jonesetal1992,Gerakinesetal1995,Whittet2005,Whittetetal2008}: (a) the  dust grains themselves (size, shape, composition, presence or absence of surface coatings), (b) B-field  direction weighted according to the distribution of dust grains along the line of sight, (c) the dust grain alignment efficiency, and (d) the physical conditions of the environment in which dust exist. Since our optical polarimetry is confined to the low-density outer parts of LDN\,1225 and due to a small dispersion in the mean B-field orientation of LDN\,1225, we interpret that the observed polarization efficiency with an index of $-$\,0.7$\pm$\,0.5 is due to the changes in the properties and alignment efficiency of dust grains as a function of extinction. 

The power-law index can also infer whether the observed polarization observations trace the B-fields in the cloud. Ideally, a power-law index of $-1$ may corresponds to ceasing of dust grain alignment producing null polarization thereby providing no clues about the B-field orientation. However, in this work, the index of $-0.7$ suggests that the dust grains are still aligned and our optical polarimetry is able to trace the B-fields at relatively less dense, outer parts of the cloud. 

\section{Summary and conclusions}\label{sec:summary_conclusions}

We have performed single $R_{C}$-band polarimetric and multi-band ($BV(RI)_{C}$) photometric observations of 
the stars distributed towards a dark globule LDN\,1225. A total of 280 stars were found to have 
$R_{C}$ polarization data satisfying the 2$\sigma$ criterion and also 689 stars are found to have 
optical plus 2MASS photometric data with uncertainties less than 0.1 mag. 
We use multi-wavelength images from DSS, WISE, {\it Spitzer}, 
and {\it Herschel}, parallaxes from GAIA DR2, and CO molecular lines data from PMO. 
In this work, we investigate the distance of LDN\,1225, distribution of 
dust and B-field orientation as a function of distance, 
importance of B-fields in the formation and evolution of cloud, 
and extinction law in the foreground and cloud mediums.

Main results of our study are summarized below:

\begin{itemize}

	\item[$\bullet$]~Based on the distance versus polarization, polarization angles, and Stokes parameters plots, 
		we ascertain the distance of LDN\,1225 as 830$\pm$83~pc.

	\item[$\bullet$]~Total sample with polarization data are classified into foreground (FG) and background (BG) 
		stars, based on the individual stellar distances in comparison to the cloud distance. 
		FG and BG stars exhibit bimodal distributions in both the level of 
		plarizations and polarization angles. 

       \item[$\bullet$]~FG stars exhibit increasing trends in their polarizations and Stokes parameters. 
	       We make use of these samples 
		to characterize the interstellar polarization (ISP) contribution and the 
		same is subtracted to infer the B-field geometry of the cloud. 

       \item[$\bullet$]~Using the dispersion in the cloud's B-field orientations, 
       gas velocity, and number density from PMO CO data, we estimate the B-field 
		strength as 56\,$\pm$\,10\,$\mu$G, by using the Davis-Chandrasekhar-Fermi relation. 
       We find that magnetic pressure is 3 times higher than the 
		turbulent pressure, and also that the mass-to-magnetic flux ratio in units of critical value 
		is less than one.
		These results imply that the dominance of B-fields over turbulence and gravity in 
		the envelope of LDN\,1225. 

	\item[$\bullet$]~The morphological correlations between B-fields and cloud geometry at 
		different optical depths, as probed by $^{12}$CO, $^{13}$CO, and C$^{18}$O 
		molecular lines, depict that B-fields might not be so important in the core scale of LDN\,1225.


	\item[$\bullet$]~Based on the {\it WISE} and {\it Herschel} images along with the CO molecular lines data, 
		we find that LDN\,1225 host two MIR and two 70\,$\mu$m sources thereby 
		reinforcing that LDN\,1225 is not a starless but star forming dark globule.

	\item[$\bullet$]~Structure function analysis suggests that the contribution from the turbulent 
	component of magnetic fields is very small compared with that of the non-turbulent 
	ordered component in LDN\,1225.

	\item[$\bullet$]~B-fields in LDN\,1225 remain coherent with the large scale B-fields of 
	CepOB3.

	\item[$\bullet$]~Using the distance versus polarizations, CO molecular lines data, and B-fields, we infer that 
		LDN\,1225 is associated and located at the same distance as CepOB3.

 	\item[$\bullet$]~Using the photometric colors, the extinction law is tested in the
                foreground and cloud mediums. We find that foreground medium is characterized with the normal extinction law,
                whereas the cloud medium is with an abnormal extinction law.

        \item[$\bullet$]~Polarization efficiency of the dust grains of LDN\,1225 decline as a function
                of extinction and yields a power-law exponent of $-$0.7\,$\pm$\,0.5
                implying that our optical polarimetry is capable of tracing B-fields
                in the low-density parts of LDN\,1225.

                In conclusion, we make use of the polarization and distance information to study the 
		dust distribution and their properties, and B-field orientation of the cloud. 
		This can serve as an important tool to probe the 3D tomography of ISM and B-fields towards 
		the molecular clouds and star forming regions. 
		Photometry has been utilized to investigate the nature of extinction law 
		to characterize the dust properties. Polarimetry along with the molecular lines data serve as an 
		efficient tool to study the correlations between the multiple polarization and cloud components.
		Multi-wavelength (dust extinction and emission) polarization data, covering different length and density 
		scales, are essential to test the star formation 
		models.

\end{itemize}

\acknowledgments

We thank the anonymous referee for his/her constructive comments which have 
improved the flow and contents of the manuscript. 
EC and SPL acknowledge support from the Ministry of Science and
Technology (MOST) of Taiwan with Grant MOST 106-2119-M-007-021-MY3.
This work is supported by National Key R\&D Program of China grant No. 2017YFA0402600 and
International Partnership Program of Chinese Academy of Sciences grant No. 114A11KYSB20160008.
EC is thankful to Prof. Wen-Ping Chen, Prof. G. C. Anupama, and Prof. G. Maheswar for the help, support, encouragement, and fruitful discussions.
EC thank Dr. Brajesh Kumar for his careful reading and helpful inputs on the draft. 

\facilities{ARIES:1m (AIMPOL polarimeter), LO:1m (TRIPOL polarimeter), IUCAA:2m (IFOSC in polarimetric mode - IMPOL), HCT:2m (HFOSC)}


{}

\appendix

\section{Color-color diagrams and estimation of total-to-selective extinction ($R_{V}$)}\label{subsec:rv_estimation}

\begin{figure*}
        \centering
\resizebox{12.5cm}{12.5cm}{\includegraphics{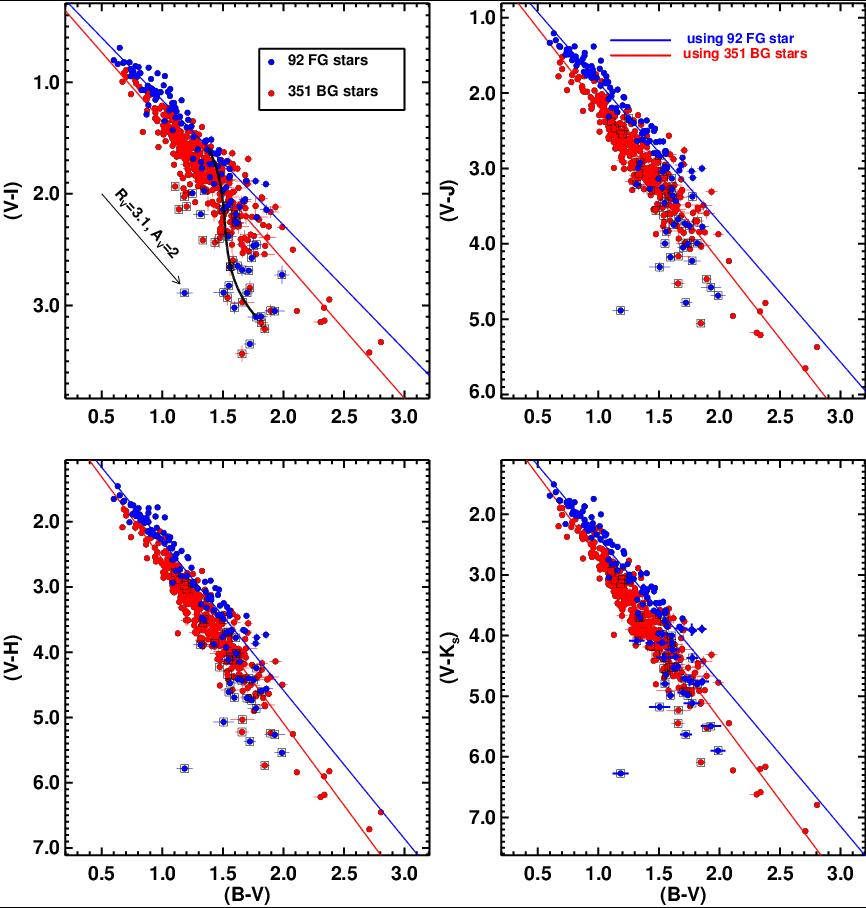}}
	\caption{The $(V-I)$, $(V-J)$, $(V-H)$, $(V-K_{s})$ versus $(B-V)$ two-color diagrams of group I stars having distances and photometric data (with uncertainties~$<$~0.1~mag in $BVIJHK_{s}$-bands). The blue and red filled circles correspond to the photometric colors of 92 FG and 351 BG stars, respectively. The blue and red lines corresponds to the weighted linear fits made on the color-color slopes of the FG and BG stars, respectively. The curve in $(V-I)$ vs $(B-V)$ represents the locus of the M-type dwarfs \citep{PetersonClemens1998}. M-type 27 FG and 16 BG dwarfs, shown with squares, are not used in the fits. The reddening vector representing a normal reddening law ($R_{V}$~$=$~3.1) is drawn with $A_{V}$~$=$~2 mag in the top left panel. \label{fig:TCD_rv_estimation1}}
\end{figure*}

\begin{figure*}
        \centering
\resizebox{12.5cm}{12.5cm}{\includegraphics{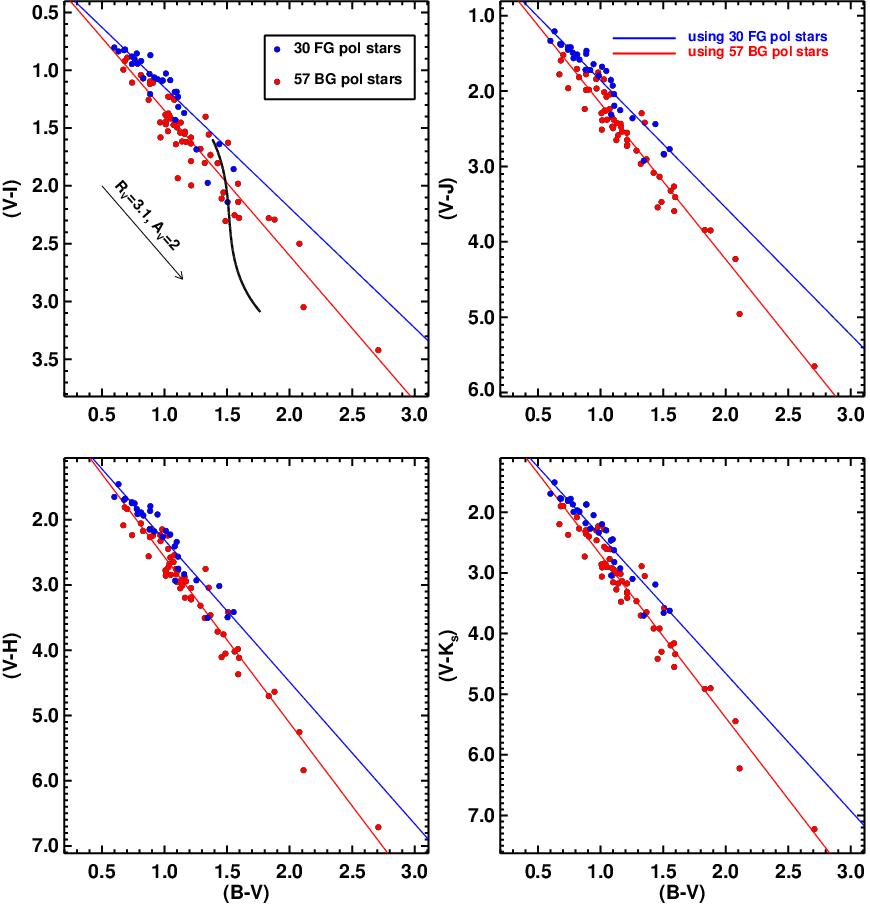}}
	\caption{$(V-I)$, $(V-J)$, $(V-H)$, $(V-K_{s})$ versus $(B-V)$ two-color diagrams of the group II stars having polarization, distance, and photometric data (uncertainties~$<$~0.1~mag in $BVIJHK_{s}$-bands). Blue and red filled circles correspond to the photometric colors of 30 FG and 57 BG stars, respectively. The blue and red lines corresponds to the weighted linear fits made on the color-color slopes of the FG and BG stars, respectively.
        The thick curve in $(V-I)$ vs $(B-V)$ represents the locus of the M-type dwarfs \citep{PetersonClemens1998}. The reddening vector representing a normal reddening law ($R_{V}$~$=$~3.1) is drawn with $A_{V}$~$=$~2 mag in the top left panel.\label{fig:TCD_rv_estimation2}}
\end{figure*}

Figure \ref{fig:TCD_rv_estimation1} shows the color-color diagrams of the 111 FG stars (with distances $<$~830~pc)
and 375 BG stars (distances between 830~pc~--~2.65 kpc) depicted using blue and red filled circles, respectively.
There exist 27 and 16 M-type stars among the FG and BG stars, respectively, and are shown with overlapping squares.
These M-type stars are identified based on a comparison of their $(B-V)$ vs $(V-I)$ colors with the intrinsic locus of M-type dwarfs \citep{PetersonClemens1998} shown with a curve in the top left panel of Figure \ref{fig:TCD_rv_estimation1}. We have excluded M-type stars from the further analysis because they would effect the true reddening law of the cloud by occupying the location of reddened background stars. The remaining 92 FG stars and 351 BG stars with photometry (hereafter group I) were used to perform the weighted linear fits and the resultant slopes are given in Table \ref{tab:colors_slopes_rv_estimation}. Similarly, shown in Figure \ref{fig:TCD_rv_estimation2}, we also estimate the slopes based on the weighted linear fits to the color-color combinations of 30 FG and 57 BG stars with photometry plus polarimetry (hereafter group II). The fitted slopes along with their uncertainties are given in Table \ref{tab:colors_slopes_rv_estimation}.
In the group I, M-type stars seem to have more contamination on the reddened BG stars
(Figure \ref{fig:TCD_rv_estimation1}), in contrast this is not pronounced
in group II (Figure \ref{fig:TCD_rv_estimation2}).

To estimate the value of $R_{V}$ for LDN\,1225, we use the following relation of the form \citep[see][]{NeckelChini1981}
\begin{equation}\label{eq:rv_esti}
R_{V}=\frac{m}{m_{normal}} R_{normal},
\end{equation}
where $m$ and $m_{normal}$ are the slopes of the two-color combinations, respectively, for the FG or BG stars and unreddened main-sequence stars following the normal extinction law (taken from the stellar models by \citealt{Bertellietal1994}, and also see Table 3 of \citealt{Pandeyetal2003}). $R_{normal}$ corresponds to normal extinction law and is considered to be 3.1.
Using the equation (\ref{eq:rv_esti}) and the fitted slopes (cf. columns 3 and 5 of Table \ref{tab:colors_slopes_rv_estimation}), we estimate $R_{V}$ values along with their uncertainties for the FG and BG stars of the two groups and are given in columns 4 and 6 of Table \ref{tab:colors_slopes_rv_estimation}. The weighted mean $R_{V}$ values are 3.10\,$\pm$\,0.01 (standard deviation = 0.14) and 3.36\,$\pm$\,0.01 (standard deviation\,$=$\,0.14), respectively, for FG and BG stars of the group I. Similarly, the weighted mean $R_{V}$ values are 2.87\,$\pm$\,0.01 (standard deviation $=$ 0.12) and 3.41\,$\pm$\,0.01 (standard deviation = 0.15), respectively, for the FG and BG stars of the group II. Discussion based on the derived $R_{V}$ values is given in Section \ref{subsec:extlaw}.

\section{Estimation of visual extinction ($A_{V}$)}\label{subsec:estim_av}

\begin{figure*}
        \centering
\resizebox{8cm}{8cm}
{\includegraphics{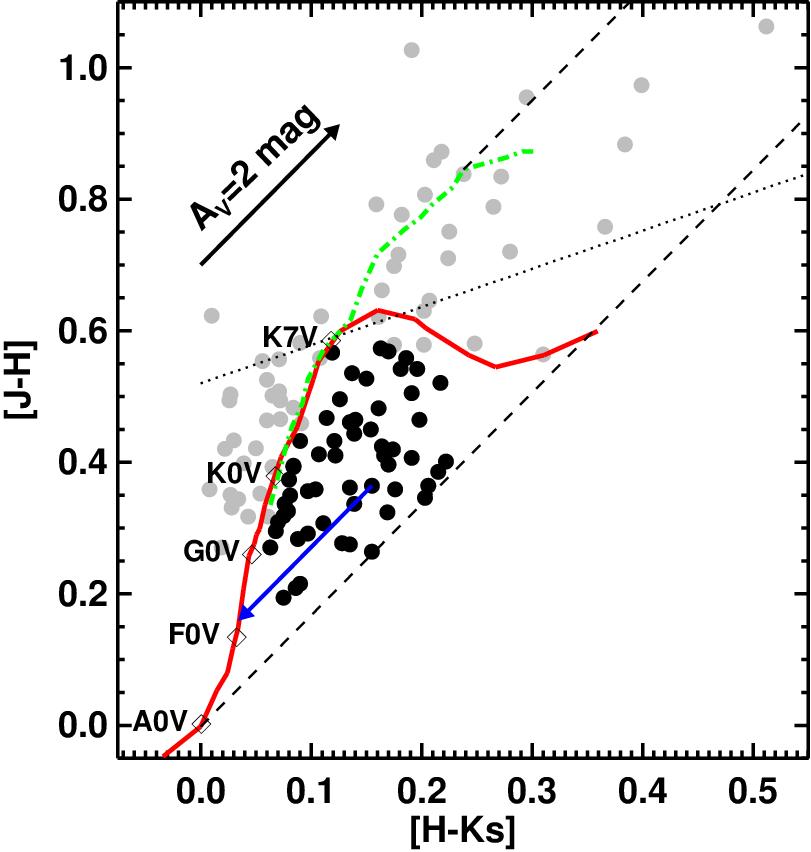}}
        \caption{NIR color-color diagram of the 113 stars having both polarization and NIR colors. Of these, 57 stars shown with filled black circles have $(J-K_{s})$~$\leq$~0.75 mag and are used to estimate the extinction ($A_{V}$) values by dereddening their observed $(J-H)$ and $(H-K_{s})$ colors. Rest of the 56 stars, shown with filled gray circles, are either distributed left side to the locus of the unreddened MS stars and giants, and the stars having $(J-K_{s})$~$>$~0.75 mag (M-type stars or giants). These stars were not used for estimating $A_{V}$ values. Blue vector depicts the extinction value of $A_{V}$~$=$~1.92 mag derived for a star with an ID 89 (cf., Table \ref{tab:av_values}), whose colors are dereddened using the NIR extinction method. The red curve and green dot-dashed lines represent the unreddened MS and giant branch \citep{BessellBrett1988}, respectively. Location of the stars with different spectral types are identified on the MS locus. The dotted line indicates the locus of unreddened CTTSs. The parallel dashed lines are the reddening vectors drawn from the tip (spectral type M4) of the giant branch (left reddening line) and from the base (spectral type A0) of the MS branch. A reference reddening vector with $A_{V}$~$=$~2 mag is shown in the upper left portion.
\label{fig:nir_ccd}}
\end{figure*}
\begin{figure*}
        \centering
\resizebox{18.0cm}{9.0cm}{\includegraphics{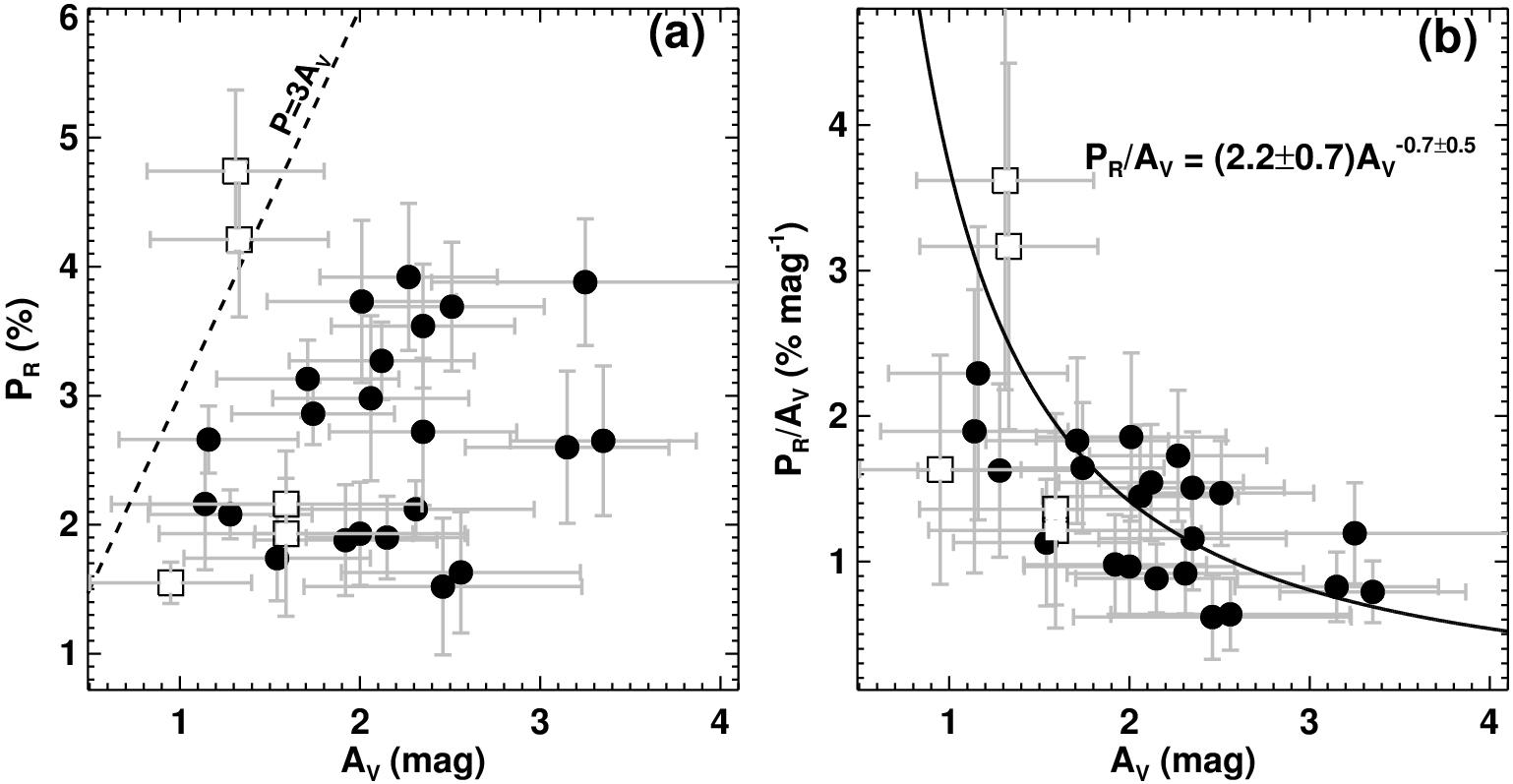}}
        \caption{(a) Extinction ($A_{V}$) versus polarization ($P_{R}$) plot for the 27 stars
        located towards LDN\,1225 and satisfying the criteria $(J-K_{s})$~$\leq$~0.75 mag
        and $A_{V}/\sigma_{A_{V}}$~$>$~2 (see Appendix~\ref{subsec:estim_av}). Dashed line denotes
        the observed polarization upper limit $P/A_{V}$~$=$~3 (\% mag$^{-1}$) \citep{Serkowskietal1975}
        corresponds to the optimum dust grain alignment in the general ISM.
(b) Same as (a) but for polarization efficiency diagram, the $A_{V}$ versus $P_{R}/A_{V}$.
	The 22 BG stars with distances $>$ 830~pc, represented with filled circles, are used in the
        weighted power-law fit as shown with a thick curve. The resultant fit values are quoted in
	the plot. It is to be noted here that, of the 27, 3 FG stars having distances $<$ 830~pc and $P_{R}$~$<$~2.5\%, and 2 BG stars
        distributed above the line $P$~$=$~$3\times A_{V}$, as depicted with open squares
        (in both panels (a) and (b)), are not used in the weighted power-law fit performed in panel (b).
\label{fig:polefficiency}}
\end{figure*}

Figure \ref{fig:nir_ccd} shows the NIR color-color ($(J-H)$ versus $(H-K_{s})$) diagram of 113 stars having polarization data. NIR photometric data of these stars were extracted from Two Micron All Sky Survey (2MASS) point source catalog \citep{Cutrietal2003}. All the sources have uncertainties less than 0.1 mag corresponds to signal-to-noise ratio (SNR) $>$ 10 and with photometric quality flag of ``AAA" in $JHK_{s}$-bands. NIR extinction method estimates $A_{V}$ value of a star by dereddening its observed colors ($(J-H)$ and $(H-K_{s})$) to match with its intrinsic colors ($(J-H)_{0}$ and $(H-K_{s})_{0}$) by using the following relations \citep[cf.][]{Maheswaretal2010}:
\begin{eqnarray}\label{eqns:nircolorsdereddened}
(J-H)_{0} = (J-H) - 0.107 \times A_{V}, \\
(H-K_{s})_{0} = (H-K_{s}) - 0.063 \times A_{V}
\end{eqnarray}
These relations utilize the extinction law of \citet{RiekeLebofsky1985}. In this method, various $A_{V}$ values having the interval of 0.01 mag were used for dereddening the observed colors, and these sets of dereddened colors were matched to the intrinsic colors as illustrated in the Figure \ref{fig:nir_ccd} using a blue arrow. The $A_{V}$ value at which the dereddened colors matched to the intrinsic colors, giving a minimum Chi-square value, yields a best $A_{V}$ of a star. NIR extinction method \citep{Maheswaretal2010} has been employed only to the 57 stars having $(J-K_{s})$~$\leq$~0.75 mag as shown with filled circles. The derived maximum value of extinction through this method comes out to be $A_{V}$~$\approx$~4 mag.
The criterion $(J-K_{s})$~$\leq$~0.75 excludes the unreddened/reddened M-type stars, giants, and classical T-Tauri stars. We also exclude stars distributed left side to the MS locus. All the excluded 56 stars are denoted with gray filled circles.
The uncertainty in $A_{V}$ values are estimated by using equation (10) of \citet{Maheswaretal2010}.

Estimated $A_{V}$ values along with uncertainties for 57 stars are listed in Table \ref{tab:av_values}.
This table also lists the IDs, coordinates, NIR colors, polarization values.
Further analysis uses only 32 stars and satisfying the criteria $A_{V}/\sigma_{A_{V}}$~$\geq$~2.
The derived $2\sigma$ $A_{V}$ values range from 0.9 mag to 3.3 mag with mean of 2.0$\pm$0.6 mag\footnote{The corresponding mean column density is 
$N_{H_{2}}$~$=$~(2.2$\pm$0.2)$\times$10$^{21}$ cm$^{-2}$ (using the relations $N_{H}$~$=$~$2N_{H_{2}}$ and 
$N_{H}$~$=$~(2.21$\pm$0.09)$\times$10$^{21}$~$A_{V}$ mag$^{-1}$ cm$^{-2}$; see \citealt{GuverOzelz2009,ValencicSmith2015}).
This value, within the error, is consistent with the column density, (3.4$\pm$1.3)$\times$10$^{21}$ cm$^{-2}$,
derived using the $^{12}$CO integrated emission of LDN\,1225 (cf., Section \ref{subsec:gas_disp_numb_dense}).}. 
Based on the $A_{V}$ and polarization values, polarization efficiency of the dust grains of LDN\,1225~ is discussed in Section \ref{subsec:ext_vs_pol_poleffi}. 


\begin{table*}
\centering
\caption{$R_{c}$-band measurements of polarized standard stars.}\label{tab:polstandresults}
\begin{tabular}{ccccc}\hline 
	Date  & $P_{R_{c}} (\%)$  & $\theta_{R_{c}} (\degr)$  &   $P_{R_{c}} (\%)$  & $\theta_{R_{c}} (\degr)$  \\
	\multicolumn{5}{c}{\hspace{2.5cm}Our work\hspace{1.6cm}\citet{Schmidtetal1992}}\\
\hline
\multicolumn{5}{c}{Polarized standard HD\,19820}\\
02 Nov 2013 & 4.6$\pm$0.1 & 114.6$\pm$0.6 & 4.53$\pm$0.03 & 114.5$\pm$0.2\\
12 Nov 2010  & 4.6$\pm$0.1 & 114.5$\pm$0.4 & 4.53$\pm$0.03 & 114.5$\pm$0.2\\
13 Dec 2010 & 4.6$\pm$0.1 & 114.2$\pm$0.4 & 4.53$\pm$0.03 & 114.5$\pm$0.2\\
 \hline
 \multicolumn{5}{c}{Polarized standard HD\,204827}\\
13 Nov 2010 & 4.9$\pm$0.1 & 59.2$\pm$0.5 & 4.89$\pm$0.03 & 59.1$\pm$0.2 \\
\hline
 \multicolumn{5}{c}{Polarized standard BD+59$\degr$\,389}\\
14 Dec 2010 & 6.5$\pm$0.1 & 97.7$\pm$0.4 & 6.43$\pm$0.02 &  98.1$\pm$0.1 \\
01 Nov 2013 & 6.3$\pm$0.1 & 98.0$\pm$0.4 & 6.43$\pm$0.02 &  98.1$\pm$0.1\\
\hline
\multicolumn{5}{c}{Unpolarized standard HD\,21447}\\
14 Dec 2010  & 0.05$\pm$0.1 & 156$\pm$56  & 0.05$\pm$0.02$^\dagger$  &  171.5$^\dagger$ \\
\hline
\end{tabular}\\
	\tablecomments{$^\dagger$ $V$-band results from \citet{Schmidtetal1992}}.
\end{table*}
\begin{table*}
	\centering
	\scriptsize
	\caption{$R_{C}$-band polarization data of 280 stars along with their photometric data from 2MASS catalog, distances from GAIA DR2, and stellar classification based on our analyses.}
	\label{tab:pol280stars}
	\begin{tabular}{lccccccccc} 
		\hline
		ID & R.A (J2000) & Dec (J2000) & $P_{R}$ & $\theta_{R}$ & $J$ & $H$ & $K_{s}$ & distance & classification\\
		   & (h:m:s) & ($\deg:\arcmin:\arcsec$) & (\%) & ($\deg$)  & (mag) & (mag) & (mag) & (pc) &  \\
		(1) & (2)   & (3)        & (4)  &  (5)      & (6)   & (7)   & (8)   & (9)  & (10) \\ 
		\hline
                   1   & 23:10:38.082   & 61:30:40.208   &   2.5 $\pm$  0.9   &    74 $\pm$  10   &   14.22 $\pm$  0.04   &   13.70 $\pm$  0.04   &   13.51 $\pm$  0.04   &     1098 $\pm$    53   &      BG          \\
   2   & 23:10:38.271   & 61:29:25.339   &   2.5 $\pm$  0.4   &    98 $\pm$   5   &   11.83 $\pm$  0.03   &   11.02 $\pm$  0.03   &   10.84 $\pm$  0.02   &     2380 $\pm$   152   &      BG          \\
   3   & 23:10:40.539   & 61:28:48.878   &   2.1 $\pm$  0.3   &    78 $\pm$   5   &   12.73 $\pm$  0.03   &   12.40 $\pm$  0.03   &   12.29 $\pm$  0.02   &      669 $\pm$    10   &      FG          \\
   4   & 23:10:41.364   & 61:30:57.006   &   1.9 $\pm$  0.2   &    51 $\pm$   3   &   12.06 $\pm$  0.03   &   11.67 $\pm$  0.03   &   11.59 $\pm$  0.02   &      460 $\pm$     4   &      FG          \\
   5   & 23:10:43.502   & 61:31:52.251   &   1.8 $\pm$  0.5   &    73 $\pm$   8   &   13.47 $\pm$  0.03   &   13.09 $\pm$  0.03   &   12.97 $\pm$  0.03   &      843 $\pm$    20   &      BG          \\
   6   & 23:10:46.677   & 61:31:29.056   &   1.8 $\pm$  0.2   &    93 $\pm$   3   &   10.84 $\pm$  0.03   &   10.01 $\pm$  0.03   &    9.78 $\pm$  0.02   &     2668 $\pm$   159   &      BG-NGC7538  \\
   7   & 23:10:47.868   & 61:31:10.794   &   1.1 $\pm$  0.3   &    51 $\pm$   7   &   12.90 $\pm$  0.03   &   12.41 $\pm$  0.03   &   12.35 $\pm$  0.03   &      517 $\pm$     6   &      FG          \\
   8   & 23:10:48.248   & 61:29:38.274   &   2.2 $\pm$  0.5   &    76 $\pm$   6   &   13.65 $\pm$  0.04   &   13.15 $\pm$  0.05   &   13.06 $\pm$  0.04   &     2078 $\pm$   145   &      BG          \\
   9   & 23:10:48.262   & 61:27:45.871   &   2.8 $\pm$  0.5   &    94 $\pm$   5   &   13.40 $\pm$  0.03   &   13.07 $\pm$  0.04   &   12.84 $\pm$  0.03   &     1061 $\pm$    32   &      BG          \\
  10   & 23:10:50.616   & 61:32:33.244   &   3.4 $\pm$  0.4   &    68 $\pm$   3   &   10.77 $\pm$  0.03   &    9.43 $\pm$  0.03   &    9.02 $\pm$  0.02   &             --         &          --      \\
  11   & 23:10:53.417   & 61:27:59.151   &   3.4 $\pm$  0.3   &    85 $\pm$   3   &   10.60 $\pm$  0.03   &    9.31 $\pm$  0.03   &    8.96 $\pm$  0.02   &     3206 $\pm$   601   &      BG-NGC7538  \\
  12   & 23:10:54.662   & 61:32:45.974   &   2.7 $\pm$  0.7   &   111 $\pm$   7   &   14.23 $\pm$  0.03   &   13.71 $\pm$  0.03   &   13.46 $\pm$  0.04   &     3812 $\pm$   615   &      BG-NGC7538  \\
  13   & 23:10:54.854   & 61:28:26.958   &   2.7 $\pm$  0.5   &    79 $\pm$   5   &   13.57 $\pm$  0.02   &   13.24 $\pm$  0.03   &   13.13 $\pm$  0.02   &     1497 $\pm$    67   &      BG          \\
  14   & 23:10:55.633   & 61:27:21.798   &   5.6 $\pm$  0.4   &    57 $\pm$   2   &   12.10 $\pm$  0.03   &   11.54 $\pm$  0.03   &   11.29 $\pm$  0.02   &     5519 $\pm$   981   &      BG-NGC7538  \\
  15   & 23:10:56.201   & 61:31:54.080   &   1.5 $\pm$  0.5   &    89 $\pm$  10   &   14.15 $\pm$  0.03   &   13.91 $\pm$  0.05   &   13.72 $\pm$  0.05   &     2141 $\pm$   179   &      BG          \\
		\hline
\end{tabular} \\
\tablecomments{A portion of the table is given here, and the entire table will be available in the online version of the paper.}
\tablecomments{$\dagger$: The stars distributed within 8$\arcmin$ area (circle in Figure \ref{fig:polvecmap_colorcompo}(a)) containing the star-forming region NGC\,7538.}
\tablecomments{$\ddagger$: The stars with NIR excess emission.}
\end{table*}




\begin{table*}
\centering
\scriptsize
\caption{Photometric data of 689 stars with their photometric uncertainties less than 0.1 mag.}\label{tab:phot689stars}
\begin{tabular}{cccccccccc} 
\hline
ID   & R.A (J2000) & Dec (J2000) & $V$    & $(B-V)$ & $(V-R)$  & $(V-I)$  &  $J$ & $H$ & $K_{s}$ \\ 
     & (h:m:s)   & ($\degr:\arcmin:\arcsec$)   & (mag)    & (mag)  & (mag) & (mag) & (mag) & (mag) & (mag)  \\ 
(1)  &  (2)        &   (3)     &   (4)  & (5)     &  (6)   &   (7)    &   (8)  & (9)  & (10) \\
\hline
    1 & 23:10:41.320 & 61:39:19.299 &   18.349 $\pm$    0.011 &    1.382 $\pm$    0.028 &    0.871 $\pm$    0.008 &    1.693 $\pm$    0.044 &    15.47 $\pm$     0.05 &    14.89 $\pm$     0.07 &    14.66 $\pm$     0.09 \\
    2 & 23:10:41.723 & 61:36:52.675 &   16.572 $\pm$    0.004 &    1.291 $\pm$    0.008 &    0.747 $\pm$    0.003 &    1.525 $\pm$    0.010 &    14.22 $\pm$     0.03 &    13.68 $\pm$     0.04 &    13.56 $\pm$     0.05 \\
    3 & 23:10:41.759 & 61:34:45.177 &   16.364 $\pm$    0.010 &    1.121 $\pm$    0.016 &    0.686 $\pm$    0.009 &    1.347 $\pm$    0.009 &    14.15 $\pm$     0.04 &    13.64 $\pm$     0.04 &    13.56 $\pm$     0.09 \\
    4 & 23:10:42.002 & 61:36:39.916 &   17.455 $\pm$    0.006 &    1.097 $\pm$    0.012 &    0.710 $\pm$    0.005 &    1.497 $\pm$    0.022 &    15.14 $\pm$     0.04 &    14.83 $\pm$     0.06 &    14.57 $\pm$     0.08 \\
    5 & 23:10:42.061 & 61:37:36.948 &   18.338 $\pm$    0.010 &    1.284 $\pm$    0.026 &    0.875 $\pm$    0.009 &    1.861 $\pm$    0.037 &    15.48 $\pm$     0.06 &    14.71 $\pm$     0.05 &    14.73 $\pm$     0.09 \\
    6 & 23:10:42.204 & 61:37:43.968 &   17.588 $\pm$    0.005 &    1.053 $\pm$    0.012 &    0.722 $\pm$    0.006 &    1.558 $\pm$    0.022 &    15.15 $\pm$     0.07 &    14.62 $\pm$     0.07 &    14.44 $\pm$     0.08 \\
    7 & 23:10:42.624 & 61:42:52.848 &   16.808 $\pm$    0.004 &    1.108 $\pm$    0.011 &    0.782 $\pm$    0.005 &    1.568 $\pm$    0.003 &    14.30 $\pm$     0.03 &    13.80 $\pm$     0.04 &    13.64 $\pm$     0.04 \\
    8 & 23:10:42.651 & 61:36:19.713 &   16.675 $\pm$    0.004 &    1.013 $\pm$    0.007 &    0.669 $\pm$    0.004 &    1.401 $\pm$    0.011 &    14.53 $\pm$     0.03 &    14.01 $\pm$     0.04 &    13.93 $\pm$     0.04 \\
    9 & 23:10:42.981 & 61:43:32.314 &   16.859 $\pm$    0.004 &    1.098 $\pm$    0.011 &    0.764 $\pm$    0.004 &    1.524 $\pm$    0.003 &    14.48 $\pm$     0.04 &    13.92 $\pm$     0.05 &    13.77 $\pm$     0.04 \\
   10 & 23:10:43.080 & 61:47:15.529 &   18.162 $\pm$    0.006 &    1.639 $\pm$    0.040 &    1.141 $\pm$    0.005 &    2.116 $\pm$    0.006 &    14.76 $\pm$     0.04 &    14.03 $\pm$     0.05 &    13.78 $\pm$     0.05 \\
   11 & 23:10:43.417 & 61:30:41.061 &   18.966 $\pm$    0.018 &    1.867 $\pm$    0.073 &    1.137 $\pm$    0.012 &    2.391 $\pm$    0.046 &    15.52 $\pm$     0.06 &    14.56 $\pm$     0.05 &    14.52 $\pm$     0.08 \\
12(5) & 23:10:43.502 & 61:31:52.251 &   15.229 $\pm$    0.006 &    0.981 $\pm$    0.003 &    0.534 $\pm$    0.002 &    1.090 $\pm$    0.006 &    13.47 $\pm$     0.03 &    13.09 $\pm$     0.03 &    12.97 $\pm$     0.03 \\
	\hline
   \hline
\end{tabular} \\
\tablecomments{A portion of the table is given here, and the entire table will be available in the online version of the paper.}
\tablecomments{In column 1, star IDs mentioned in parentheses are the stars having polarization data (Table \ref{tab:pol280stars})}.
\end{table*}



\begin{table*}
\centering
\scriptsize
\caption{Weighted mean polarization measurements of FG and BG stars (ISP corrected) of various fields of CepOB3. Also given are the central coordinates and number of stars in each field, the passband, and the instrument or source used for the data.}\label{tab:weightedmeanPPAcepob3}
\begin{tabular}{cccccccc}\hline \hline
Field/HD number & RA (J2000) & Dec (J2000) & weighted mean $P$ & weighted mean $\theta$ & pass-band & No. of stars & instrument (or) source \\
         & (h:m:s) & ($\degr:\arcmin:\arcsec$)    & ($\%$)    & ($\degr$)    &          &         &              \\
 (1)&(2)&(3)&(4) &(5) &  (6) & (7) & (8)\\
\hline
\multicolumn{8}{c}{FG stars with d $<$ 830 pc}\\
\hline

        1 & 22:56:53.332 & 62:09:58.572    &     0.54  $\pm$  0.14     &       63   $\pm$      7   &   R           &   4  &   AIMPOL \\
        2 & 22:56:33.530 & 62:04:54.912    &     2.20  $\pm$  0.10     &       73   $\pm$      1   &   R           &   5  &    ,,   \\
        3 & 22:56:07.713 & 61:58:02.352    &     1.19  $\pm$  0.08     &      109   $\pm$      2   &   R           &   5  &    ,,   \\
        4 & 22:57:41.210 & 62:41:51.792    &     1.29  $\pm$  0.13     &       69   $\pm$      3   &   R           &   2  &   IMPOL   \\
        5 & 22:58:16.408 & 62:43:25.176    &           --              &             --            &   R           &   0  &     ,,    \\
        6 & 22:57:02.560 & 62:39:01.188    &     0.61  $\pm$  0.03     &       68   $\pm$      1   &   R           &   4  &    ,,   \\
        7 & 22:56:50.930 & 62:34:53.292    &     1.31  $\pm$  0.06     &       76   $\pm$      1   &   R           &   3  &    ,,   \\
        8 & 22:56:51.355 & 62:31:18.084    &     1.41  $\pm$  0.03     &       83   $\pm$      1   &   R           &   4  &    ,,   \\
        9 & 22:58:19.392 & 62:38:45.744    &     0.11  $\pm$  0.06     &       75   $\pm$     14   &   R           &   1  &    ,,   \\
       10 & 22:57:34.209 & 62:35:53.484    &     1.00  $\pm$  0.07     &      163   $\pm$      2   &   R           &   3  &    ,,   \\
	\hline
\hline
\end{tabular} \\
\tablecomments{RA and Dec (except those of the stars from \citet{Heiles2000}) are the central coordinates of the observed fields.}
\tablecomments{A portion of the table is given here, and the entire table will be available in the online version of the paper.}
\end{table*}


\begin{table*}
	\centering
	\caption{Spatial extents and position angles of LDN\,1225 based on the CASA 2D Gaussian function fitting to the moment 0 maps of $^{12}$CO, $^{13}$CO, and C$^{18}$O}
   \label{tab:cloudpas}
	\begin{tabular}{ccccccc} 
		\hline
\hline
		Line & R.A (J2000) & Dec (J2000) & $FWHM_{major}$  &  $FWHM_{minor}$ &  $\theta_{cloud}$   &  $\Delta_{\theta}$~$=$~$|$~$\theta_{cloud}$~$-$~$\theta_{C}$~$|$ \\
             & (h:m:s) & ($\degr$:$\arcmin$:$\arcsec$) & (arcmin) & (arcmin)  & (\degr) & ($\degr$) \\
             (1)  & (2) & (3) & (4) & (5) & (6) & (7) \\
		\hline
		$^{12}$CO  & 23:12:16.62  &  61:38:24.27  &  8.6$\pm$0.2 &  4.4$\pm$0.1  &  102$\pm$1    &    4$\pm$ 11  \\         
		$^{13}$CO  & 23:12:01.11  &  61:39:13.50  &  4.4$\pm$0.1 &  3.1$\pm$0.1  &   52$\pm$3    &   54$\pm$ 11  \\
C$^{18}$O  & 23:11:55.60  &  61:39:15.00  &  1.8$\pm$0.1 &  1.3$\pm$0.1  &   54$\pm$8    &   52$\pm$14  \\
\hline
\hline
	\end{tabular} \\
	\tablecomments{$\theta_{C}$~$=$~106$\pm$11$\degr$, the mean B-field orientation in LDN\,1225 after foreground (interstellar) polarization correction (cf. Section \ref{subsec:fg_subtraction}).}
\end{table*}


\begin{table*}
\centering
\caption{$R_{V}$ values estimated based on fitted slopes over different color-color combinations of FG and BG stars belong to two groups of stars. Group I with photometric data, while Group II with both photometric and polarimetric data.}\label{tab:colors_slopes_rv_estimation}
\begin{tabular}{ccccc}\hline \hline
color-color combination & $m\pm\sigma_{m}$ & $R_{V}\pm\sigma_{R_{V}}$ & $m\pm\sigma_{m}$ & $R_{V}\pm\sigma_{R_{V}}$ \\
(1)&(2)&(3)&(4) &(5)\\
\hline
\multicolumn{5}{c}{Group I: Stars with photometric data}\\
	\multicolumn{5}{c}{\hspace{2.5cm} 92 FG stars \hspace{2.5cm} 351 BG stars}\\
\hline
 $(V-I)/(B-V)$      & -1.12  $\pm$  0.01   &   3.15  $\pm$  0.01   &   -1.24  $\pm$  0.01   &   3.49  $\pm$  0.01   \\
 $(V-J)/(B-V)$      & -1.86  $\pm$  0.01   &   2.94  $\pm$  0.02   &   -2.06  $\pm$  0.01   &   3.26  $\pm$  0.01   \\
 $(V-H)/(B-V)$      & -2.28  $\pm$  0.01   &   2.92  $\pm$  0.02   &   -2.52  $\pm$  0.01   &   3.22  $\pm$  0.01   \\
 $(V-K_{s})/(B-V)$  & -2.38  $\pm$  0.01   &   2.84  $\pm$  0.01   &   -2.68  $\pm$  0.01   &   3.19  $\pm$  0.01   \\
	\hline
\multicolumn{5}{c}{Group II: Stars with photometric plus polarimetric data}\\
	\multicolumn{5}{c}{\hspace{2.5cm} 30 FG stars \hspace{2.5cm} 57 BG stars}\\
\hline
 $(V-I)/(B-V)$      &  -1.04  $\pm$   0.01   &   2.93  $\pm$   0.01   &   -1.26  $\pm$   0.01    &  3.54  $\pm$   0.01 \\
 $(V-J)/(B-V)$      &  -1.69  $\pm$   0.02   &   2.67  $\pm$   0.03   &   -2.07  $\pm$   0.01    &  3.27  $\pm$   0.02 \\
 $(V-H)/(B-V)$      &  -2.18  $\pm$   0.02   &   2.79  $\pm$   0.03   &   -2.54  $\pm$   0.01    &  3.26  $\pm$   0.02 \\
 $(V-K_{s})/(B-V)$  &  -2.27  $\pm$   0.02   &   2.71  $\pm$   0.02   &   -2.69  $\pm$   0.01    &  3.20  $\pm$   0.01 \\
\hline
\hline
\end{tabular} \\
\end{table*}

\begin{table*}
\centering
\scriptsize
	\caption{Total extinction $A_{V}$ values of 57 stars derived using NIR extinction method. ID numbers, coordinates, NIR colors, 
	and polarization values are also given.}\label{tab:av_values}
\begin{tabular}{ccccccc}\hline \hline
	ID  & RA (J2000) & Dec (J2000) & $[J-H]$  & $[H-K_{S}]$   & $P_{R}$  & $A_{V}$  \\
	    & ($\degr$)  & ($\degr$)   & (mag)    & (mag)    &    (\%)  &   (mag)    \\
 (1)&(2)&(3)&(4) &(5) &  (6) & (7) \\
\hline
   05 & 23:10:43.502 & 61:31:52.251 &   0.40 $\pm$   0.04 &   0.08 $\pm$   0.04 &    1.8 $\pm$    0.5 &    0.4 $\pm$    0.5 \\
   12 & 23:10:54.662 & 61:32:45.974 &   0.52 $\pm$   0.04 &   0.22 $\pm$   0.05 &    2.7 $\pm$    0.7 &    2.8 $\pm$    0.6 \\
   15 & 23:10:56.201 & 61:31:54.080 &   0.26 $\pm$   0.05 &   0.16 $\pm$   0.07 &    1.5 $\pm$    0.5 &    2.5 $\pm$    0.8 \\
   19 & 23:11:00.097 & 61:33:54.450 &   0.43 $\pm$   0.05 &   0.12 $\pm$   0.05 &    2.5 $\pm$    0.7 &    1.0 $\pm$    0.6 \\
   20 & 23:11:03.445 & 61:41:30.120 &   0.39 $\pm$   0.04 &   0.08 $\pm$   0.04 &    2.3 $\pm$    0.2 &    0.4 $\pm$    0.5 \\
   42 & 23:11:18.675 & 61:46:42.571 &   0.42 $\pm$   0.04 &   0.17 $\pm$   0.04 &    1.9 $\pm$    0.3 &    2.2 $\pm$    0.4 \\
   44 & 23:11:20.161 & 61:31:08.886 &   0.41 $\pm$   0.04 &   0.11 $\pm$   0.05 &    1.8 $\pm$    0.4 &    0.8 $\pm$    0.6 \\
   46 & 23:11:23.872 & 61:34:43.075 &   0.57 $\pm$   0.04 &   0.16 $\pm$   0.04 &    2.1 $\pm$    0.2 &    1.3 $\pm$    0.5 \\
   47 & 23:11:24.807 & 61:31:16.078 &   0.47 $\pm$   0.04 &   0.11 $\pm$   0.05 &    3.5 $\pm$    0.8 &    0.8 $\pm$    0.6 \\
   51 & 23:11:27.976 & 61:43:49.180 &   0.37 $\pm$   0.03 &   0.08 $\pm$   0.04 &    1.5 $\pm$    0.2 &    0.3 $\pm$    0.5 \\
   54 & 23:11:31.673 & 61:34:05.188 &   0.19 $\pm$   0.04 &   0.08 $\pm$   0.04 &    3.2 $\pm$    0.2 &    0.7 $\pm$    0.5 \\
   55 & 23:11:32.560 & 61:33:12.301 &   0.34 $\pm$   0.04 &   0.14 $\pm$   0.05 &    3.9 $\pm$    0.6 &    1.7 $\pm$    0.5 \\
   57 & 23:11:33.231 & 61:32:13.758 &   0.31 $\pm$   0.03 &   0.07 $\pm$   0.04 &    1.3 $\pm$    0.2 &    0.4 $\pm$    0.4 \\
   58 & 23:11:33.251 & 61:43:31.011 &   0.29 $\pm$   0.04 &   0.10 $\pm$   0.04 &    1.5 $\pm$    0.2 &    0.9 $\pm$    0.4 \\
   59 & 23:11:34.034 & 61:45:52.336 &   0.36 $\pm$   0.04 &   0.21 $\pm$   0.05 &    2.6 $\pm$    0.6 &    3.2 $\pm$    0.6 \\
\hline
\hline
\end{tabular} \\
\tablecomments{ID numbers are same as those of Table \ref{tab:pol280stars}.}
\tablecomments{A portion of the table is given here, and the entire table will be available in the online version of the paper.}
\end{table*}

\clearpage

\end{document}